\documentclass[aps,10pt,twocolumn,prb,floatfix,superscriptaddress]{revtex4-1}
\usepackage{hyperref}
\usepackage[usenames,dvipsnames]{color}
\usepackage{amsmath}
\usepackage{amssymb}
\usepackage{graphicx}
\usepackage{epstopdf}
\usepackage{tabularx}
\usepackage[disable]{todonotes}
\usepackage{tablefootnote}
\usepackage{threeparttable}
\usepackage{color}
\usepackage[normalem]{ulem}

\graphicspath{ {../make_data_plots/} }

\begin{document}
\title{Solutions of the Two Dimensional Hubbard Model: Benchmarks and Results from a Wide Range of Numerical Algorithms}

\author{J. P. F. LeBlanc}
\affiliation{Department of Physics, University of Michigan, Ann Arbor, Michigan 48109, USA} 

\author{Andrey E. Antipov}
\affiliation{Department of Physics, University of Michigan, Ann Arbor, Michigan 48109, USA} 
\author{Federico Becca}
\affiliation{CNR-IOM-Democritos National Simulation Centre and International School
for Advanced Studies (SISSA), Via Bonomea 265, I-34136, Trieste, Italy}

\author{Ireneusz W. Bulik}
\affiliation{Department of Chemistry, Rice University, Houston, TX 77005}

\author{Garnet Kin-Lic Chan}
\affiliation{Department of Chemistry, Princeton University, Princeton, NJ, 08544}

\author{Chia-Min Chung}
\affiliation{Department of Physics and Astronomy, University of California - Irvine, Irvine, CA 92617, USA}

\author{Youjin Deng}
\affiliation{Hefei National Laboratory for Physical Sciences at Microscale and Department of Modern Physics, University of Science and Technology of China, Hefei, Anhui 230026, China}

\author{Michel Ferrero}
\affiliation{Centre de Physique Th´eorique, Ecole Polytechnique, CNRS, 91128 Palaiseau Cedex, France}

\author{Thomas M. Henderson}
\affiliation{Department of Chemistry, Rice University, Houston, TX 77005}
\affiliation{Department of Physics and Astronomy, Rice University, Houston, TX 77005}

\author{Carlos A. Jim\'enez-Hoyos}
\affiliation{Department of Chemistry, Rice University, Houston, TX 77005}

\author{E. Kozik}
\affiliation{Department of Physics, King's College London, Strand, London WC2R 2LS, UK}

\author{Xuan-Wen Liu}
\affiliation{Hefei National Laboratory for Physical Sciences at Microscale and Department of Modern Physics, University of Science and Technology of China, Hefei, Anhui 230026, China}

\author{Andrew J. Millis}
\affiliation{Department of Physics, Columbia University, New York, New York 10027, USA}

\author{N. V. Prokof'ev}
\affiliation{Department of Physics, University of Massachusetts, Amherst, MA 01003, USA}
\affiliation{Russian Research Center ``Kurchatov Institute'', 123182 Moscow, Russia}

\author{Mingpu Qin}
\affiliation{Department of Physics, College of William and Mary, Williamsburg, VA 23187}

\author{Gustavo E. Scuseria}
\affiliation{Department of Chemistry, Rice University, Houston, TX 77005}
\affiliation{Department of Physics and Astronomy, Rice University, Houston, TX 77005}

\author{Hao Shi}
\affiliation{Department of Physics, College of William and Mary, Williamsburg, VA 23187}

\author{B.V. Svistunov}
\affiliation{Department of Physics, University of Massachusetts, Amherst, MA 01003, USA}
\affiliation{Russian Research Center ``Kurchatov Institute'', 123182 Moscow, Russia}

\author{Luca F. Tocchio}
\affiliation{CNR-IOM-Democritos National Simulation Centre and International School
for Advanced Studies (SISSA), Via Bonomea 265, I-34136, Trieste, Italy}

\author{I.S. Tupitsyn}
\affiliation{Department of Physics, University of Massachusetts, Amherst, MA 01003, USA}

\author{Steven R. White}
\affiliation{Department of Physics and Astronomy, University of California - Irvine, Irvine, CA 92617, USA}

\author{Shiwei Zhang}
\affiliation{Department of Physics, College of William and Mary, Williamsburg, VA 23187}

\author{Bo-Xiao Zheng}
\affiliation{Department of Chemistry, Princeton University, Princeton, NJ, 08544}

\author{Zhenyue Zhu}
\affiliation{Department of Physics and Astronomy, University of California - Irvine, Irvine, CA 92617, USA}

\author{Emanuel Gull}
\email{egull@umich.edu}
\affiliation{Department of Physics, University of Michigan, Ann Arbor, Michigan 48109, USA} 

\collaboration{The Simons Collaboration on the Many-Electron Problem}

\date{\today}

\begin{abstract}
Numerical results for ground state and excited state properties (energies, double occupancies, and Matsubara-axis self energies) of the single-orbital Hubbard model on a two-dimensional square lattice are presented, in order to provide an assessment of our ability to compute accurate results in the thermodynamic limit.  Many methods are employed, including auxiliary field quantum Monte Carlo, bare and bold-line diagrammatic Monte Carlo, method of dual fermions, density matrix embedding theory, density matrix renormalization group, dynamical cluster approximation, diffusion Monte Carlo within a fixed node approximation, unrestricted coupled cluster theory, and multi-reference projected Hartree-Fock. Comparison of results obtained by different methods allows for the identification of uncertainties and systematic errors. The importance of extrapolation to converged thermodynamic limit values is emphasized. Cases where agreement between different methods is obtained establish benchmark results that may be useful in the validation of new approaches and the improvement of existing methods.
\end{abstract}

\maketitle


\section{Introduction}
The ``many-electron problem'' of providing a useful and sufficiently accurate calculation of the properties of systems of large numbers of interacting electrons is one of the grand scientific challenges of the present day. Improved solutions are needed both for the practical problems of materials science and chemistry and for the basic science questions of determining the qualitative behaviors of interacting quantum systems. 

While many problems of implementation arise, including calculation of the multiplicity of orbitals and interaction matrix elements needed to characterize real materials, the fundamental difficulties are that the dimension of the Hilbert space needed to describe an interacting electron system grows exponentially in the system size, so that direct diagonalization is not practical except for small systems, and that  the minus sign associated with the Fermi statistics of electrons leads to exponentially slow convergence of straightforward Monte Carlo calculations. It is generally accepted that a complete solution to the many-electron problem cannot be obtained in polynomial time.

The difficulties associated with obtaining a complete solution have motivated the development over the years of approximate methods, and comparison of the different approximations remains a crucial open question. In this paper we address this issue in the context of  the repulsive-interaction Hubbard model \cite{Hubbard63,gutzwiller:1963,Kanamori63} defined on a two-dimensional square lattice. The Hubbard model  is one of the simplest models of interacting fermions, but despite its simplicity exhibits a wide range of correlated electron behavior including  interaction-driven metal-insulator transitions, superconductivity, and magnetism. The precise behavior depends delicately on parameters, creating an interesting challenge for theory and computation. 

Exact solutions are available for one-dimensional\cite{Lieb68} and infinite-dimensional cases.\cite{,metzner:1989,MH89} High temperature series expansions provide numerically exact results, but only for temperatures too high to be relevant for physically interesting situations.\cite{Oitmaa06} In general dimensions at relevant temperatures, only approximate solutions are available. 
In some cases, these provide rigorous bounds bounds to the ground state energies or other thermodynamic properties.\cite{valent:1991,valent:1991:2}
Analytical perturbative methods can provide information about the behavior at very small interaction strength \cite{kagan:1989,baranov:1992,chubukov:1992,chubukov:1993,hlubina:1999,zanchi:1996,Halboth00}  and  at very large interaction strength (for the special case of nearly one electron per site),\cite{anderson:1959,Nagaoka66} but outside of these limits obtaining results requires numerical methods.\cite{scalapino:2007} Other techniques such as diagrammatic resummation are expected to work well in the weak coupling regime.\cite{gukelberger:2015}
  The known numerical methods are based on approximation schemes. Among the approximations employed are  study of  finite systems (either directly or via embedding constructions), use of  variational wave functions and evaluation of  subsets of all possible Feynman diagrams.  Controlling these approximations and assessing the remaining uncertainties is a challenging but essential task, requiring analysis of results obtained from different methods.  The past decade has seen the development of interesting new methods and substantial improvements in capabilities of previously developed approaches, suggesting that the time is ripe for a careful comparison.

In this paper we undertake this needed task. Our aim is to assess the state of our knowledge of the Hubbard model, identifying parameter regimes where reliable results have been established and regimes where further work is needed. In regimes where reliable results have been established our results will serve as benchmarks to aid in the development and  validation of new methods.  This improved understanding of the Hubbard model will serve as a tool to analyze  methods for solving the general many electron problem.

We take the view that the only meaningful points of comparison between methods are results for the actual thermodynamic limit of the Hubbard model, with the uncertainties arising from the needed extrapolations (to infinite system size, to all diagrams, to infinite statistical precision in Monte Carlo calculations etc) quantified.  However, examination of results obtained at different stages of the extrapolation sequence for a given method provide considerable information. Therefore, we present, when needed, both converged values and the intermediate results from which these were obtained.    

 The methods considered are auxiliary field quantum Monte Carlo (AFQMC),\cite{zhang:2013,shi:tbd, shi:2013} bare and bold-line diagrammatic Monte Carlo (DiagMC),\cite{prokofiev:1998, Mishchenko:2000, vanhoucke:2010} the dual fermion method (DF),\cite{Rubtsov2008} density matrix embedding theory (DMET),~\cite{Knizia2012,Knizia2013} density matrix renormalization group theory (DMRG),\cite{white:1992,white:1993} cluster dynamical mean field theory in the dynamical cluster approximation (DCA),\cite{maier:2005} diffusion Monte Carlo based on a fixed node approximation (FN),\cite{gros:1988,gros:1989,yokoyama:1987,yokoyama:1987:2,tocchio:2008,trivedi:1990,tenhaaf:1995} unrestricted coupled cluster theory including singles and doubles (UCCSD), and
in certain cases, higher excitations,\cite{Paldus1999,Bartlett2007,ShavittBartlett} and multi-reference projected Hartree-Fock (MRPHF).\cite{rodr:2012,rodr:2013}

We examine energies and double occupancies, which are single numbers and can be obtained by essentially all methods, enabling  straightforward comparison.  We also consider properties related to the electron Green's function, which at this stage are only available from a few methods. The results obtained from different techniques enables us to identify regimes of phase space that are well understood, in the sense that several different methods provide results that are converged and agree within (reasonable) errors, and regimes that are not  well understood, in the sense that there is as yet no agreement between different methods. We show excellent agreement and small uncertainties between numerically exact techniques at half filling (all coupling strengths), weak coupling (all carrier concentrations) and for carrier concentrations far from half filling (most interaction strengths).  For carrier concentrations near to half filling and for intermediate interaction strength, results can be obtained, but the resulting uncertainties are much larger in general, and more difficult to eliminate.  We surmise that at least part of this uncertainty has a physical origin related to the presence of several competing phases, leading to sensitive dependence on parameters. 

The rest of this paper is organized as follows. In section \ref{Model} we define the model, delineate parameter regimes and define and discuss the observables of interest in this paper. In section \ref{methods} we summarize the methods, giving brief descriptions and focusing on issues most relevant to this paper while referring the reader to the literature for detailed descriptions. Section \ref{sec:extrapol} demonstrates the importance of the extrapolation of results to the thermodynamic limit and discusses the issues involved in the extrapolations.  In sections \ref{results_intermediate_strong}, \ref{results_intermediate} and \ref{results_weak} we present static observables in the strong coupling, intermediate coupling, and weak coupling regimes respectively.   Section \ref{results_kw_plots} presents momentum and frequency dependence at finite $T$.  A conclusion summarizes the work, outlines the important areas where our present-day knowledge is inadequate, and indicates directions for future research. Supplementary material presents the thermodynamic limit values obtained here.  A database of results  is made available electronically.\cite{GithubBenchmarkRepo,suppl}

\section{The Hubbard Model \label{Model}}
The Hubbard model is defined by the Hamiltonian
\begin{equation}\label{eqn:H}
H=-\sum\limits_{ i,j, \sigma}t_{ij}  \left(c_{i\sigma}^\dagger c_{j\sigma}  + H.c.\right) +U\sum\limits_i n_{i\uparrow}n_{i\downarrow},
\end{equation}
where $c^\dagger_{i\sigma}$ ($c_{i\sigma}$) creates (annihilates) an electron with spin $\sigma=\uparrow$,$\downarrow$ on site $i$,  $n_{i\sigma}=c^\dagger_{i\sigma}c_{i\sigma}$ is the number operator, and $t_{ij}$ denotes the hopping term.  In this work, we will restrict the discussion to the repulsive Hubbard model ($U>0$) defined on a two dimensional square lattice; we further assume that the hopping contains only nearest neighbor ($t_{ij}=t$) and second nearest neighbor ($t_{ij}=t^\prime$) terms.  $t$ is used to set the scale of all energies presented in this work.

We consider interaction strengths ranging from $U/t=2$ to $U/t=12$ and focus on temperatures where high-temperature expansion methods fail.\cite{Oitmaa06} We  give representative results for the ground state and temperatures  $T/t = 0.125$, $0.25$, and $0.5$.  Table~\ref{table:parms} contains a complete list of the parameters studied.
\begin{table}
\begin{tabular}{|p{3cm}|*{5}{|r}|}
\hline
Parameters     &  \multicolumn{5}{|c|}{Parameters studied} \\
\hline
$t'/t$            & -0.2 & 0 & 0.2 &  &      \\
$U/t$           & 2 & 4 & 6 & 8 &  12       \\
$n$    & 1.0 & 0.875 & 0.8 & 0.6 & 0.3\\
$T/t$     & 0 & 1/8 & 1/4 & 1/2 &   \\\hline
\end{tabular}
\caption{\label{table:parms}  Parameters studied for the Hubbard model. $t$ denotes the nearest-neighbor hopping, $t'$ the next nearest neighbor hopping, $U$ the interaction strength, $n$ the density, and $T$ the temperature.}
\end{table}

At zero temperature, we systematically compute the energy per site and the double occupancy, and we present also some data on the nature of the order and the order parameter, where an ordered phase is found. At non-zero temperature, we also present dynamical information, in particular  the Matsubara self-energy. We focus on values at a given density, rather than at a given chemical potential. This implies that methods based on a grand-canonical formulation need to adjust the chemical potential to find the right density, leading to  additional uncertainty in computed quantities.

\section{Methods \label{methods}} 
\subsection{Overview}
Many numerical methods provide solutions to the Hubbard Hamiltonian on a finite size lattice.  In this work, we restrict attention to techniques which can access systems large enough that an extrapolation to the infinite system can be performed, and our aim is to obtain results for the thermodynamic limit. Also important to our analysis is the assessment of uncertainties, either by providing an unbiased error bar or an error bar that contains all errors {\it except} for a systematic contribution which may be assessed by comparison to other methods or reference systems. 

We consider three broad classes of methods: ground state wave function methods, embedding methods, and Green's function methods. The distinction between these methods is not sharp; several of the algorithms fit into multiple categories, but the categorization provides a useful way to organize a discussion of the different kinds of uncertainties. 

Wave function methods construct an approximation to the ground state wave function of a system. Expectation values of observables (energies and correlation functions) are then evaluated by applying operators to this ground state wave function. The issues are the accuracy of the wave function for a given system size, and the accuracy of the extrapolation to the thermodynamic limit. AFQMC (with and without constrained path), UCCSD, FN, MRPHF, and DMRG are wave function methods.

Embedding methods approximate properties of the full system (for example the self energy or density matrix) by the solution of a finite cluster self-consistently embedded in an appropriately designed infinite lattice. The full solution of the original problem is recovered as the cluster size is taken to infinity. Errors in  embedding methods arise from three sources: the solution of the cluster problem, the convergence of the self-consistency loop that performs the embedding, and finite cluster size, with maximum cluster sizes depending on the method and ranging from $16$ to $100$.  The DMET, DCA, DF   are embedding methods. 

Green's function methods are defined here as methods based on stochastic evaluations of many-body perturbation series. They provide many-body self-energies, $\Sigma(k,i\omega_n)$,  and Green's functions, $G(k, i \omega_n)$, as functions of momenta, $k$, and fermionic matsubara frequency, $i\omega_n$, from which other observables (energies and  densities) can be calculated. 
Techniques which produce real-frequency (rather than matsubara frequency) information\cite{bulla:2008,dagotto:1994,vanschilfgaarde:2006,louie:1986} are either restricted to small systems, molecules or impurity models or work best at weak coupling.
DiagMC is a Green's function method  formulated directly in the infinite lattice; the main issues for this method are the accuracy of the stochastic approximation to the full diagrammatic expansion and the systematic truncation and extrapolation of the series. The DF and DCA techniques use Green's function techniques to evaluate the impurity problem and the expansion around it; they therefore are subject both to embedding uncertainties and to the uncertainties arising from the evaluation of the diagrammatic expansion. 

\subsection{Auxiliary-field Quantum Monte Carlo(AFQMC)}
Auxiliary-field Quantum Monte Carlo (AFQMC) is a ground-state wave-function method based on the idea that in the limit $\beta\rightarrow\infty$ the operator $e^{-\beta H}$ applied to an initial wave function $|\psi^{(\beta=0)}\rangle$ projects out the ground state of the Hamiltonian $H$. The projection is formulated as an imaginary-time path integral that is  stochastically evaluated with
the help of auxiliary fields introduced by a Hubbard-Stratonovich transformation.  The method is applied to finite size lattices and an extrapolation to the infinite lattice case is required. If the calculation is converged, the exact ground state energy and wave function for the lattice are obtained. The issues are the convergence of the stochastic evaluation of the projector and, when particle-hole symmetry is broken,   the presence of a sign problem. The sign problem is managed using  a constrained path approximation, which introduces a potential systematic error that must be quantified by comparison to other methods. For an introduction to the basics of AFQMC methods, see, {\it e.g.}, ~Ref.~\onlinecite{zhang:2013}.

In this manuscript we present results obtained from two ground-state auxiliary-field quantum Monte Carlo (AFQMC) methods. The first \cite{shi:tbd} is based on the ground-state path integral form of AFQMC,\cite{blankenbecler:1981, sugiyama:1986,white:1989} but introduces several new algorithmic ingredients including an acceleration technique\cite{shi:2015} (with force bias \cite{zhang:2013,zhang:2003}) in the Metropolis sampling and control of Monte Carlo variance divergence.\cite{shi:tbd}
This approach is applied to systems at half filling with $t'=0$, where the sign problem\cite{loh:1990} is absent because of particle-hole symmetry.\cite{hirsch:1985} The algorithmic improvements allow us to reach longer imaginary time in the calculations, achieve a higher acceptance ratio, and greatly reduce the Monte Carlo variance.\cite{shi:tbd} The second approach we employ, to treat cases where the sign problem does occur, is referred to as the  constrained path Monte Carlo method.\cite{zhang:1997, nguyen:2014} This approach controls the sign problem with a constraint (implemented via a choice of trial wave function\cite{zhang:2013}  $|\Psi_T\rangle$) on the paths in auxiliary-field space, which allows stable calculations for arbitrarily long imaginary time and system size. 


Both methods obtain the ground state of the Hamiltonian for a supercell of size $L\times L$ under twist-averaged boundary conditions.\cite{lin:2001,chang:2008}
 The ground state is obtained by use of  Monte Carlo methods to estimate $|\psi^{(\beta)}\rangle = e^{-\beta H}\psi^{(\beta=0)}\rangle$. The total projection length is typically $\beta = 64$ in the ground state projection method, although test calculations with  imaginary-time lengths several times larger were performed. The convergence error from finite values of $\beta$ is negligible. In the constrained path method, the runs are open-ended and tend to correspond to much larger values of $\beta$.   In both the ground state projection and constrained path methods  the statistical error from the Monte Carlo calculation can be reliably estimated (one-standard deviation error bars reported). The systematic error from the constrained path method is
not variational, but depends on $|\Psi_T\rangle$ in the sense that it vanishes if $|\Psi_T\rangle$ is exact. Its magnitude will be quantified below by comparison to other techniques.

A Trotter decomposition is used in the imaginary time evolution. The Trotter error from the finite time step, $\Delta\tau=\beta/n$, must be extrapolated to zero. This extrapolation can be controlled and does not make a significant contribution to the error budget.  Most calculations reported here use a time-step fixed at $\Delta \tau = 0.01$ in units of $t$. 

Results obtained for finite $L$ are averaged over twist-angle, $\Theta$,  to remove one-body finite size effects. For small systems a large number of $\Theta$ values are needed ($\sim 200$ for $4 \times 4$ or $6 \times 6$),\cite{nguyen:2014,chang:2008, chang:2010} but for larger systems, far  fewer $\Theta$ are required to reach the same level of accuracy (for a  $L\times L$ system with $L=20$, averaging over $5$ twist angles is sufficient). 
These results are then extrapolated to $L\rightarrow\infty$. The extrapolation requires care because the ground state depends on the system geometry. For $n=0.875$, we used rectangular supercells (mostly $8\times32$, checked with sizes $8 \times 16$, $16 \times 16$ and $8 \times 48$  for consistency) to accommodate spin- and charge-density wave orders. The extrapolation also requires careful attention to the functional form of the leading finite volume correction.\cite{huse:1988, neuberger:1989, fisher:1989,chang:2010, white:1998} Our final results at the thermodynamic limit include all statistical errors, and a conservative estimate of the uncertainty resulting from the extrapolation of $L \to \infty$ in order to remove the two-body finite-size effects (The fit includes $1/L^3$ and $1/L^4$ terms\cite{chang:2008,chang:2010,huse:1988} for the energies, and $1/L$ for magnetization $m^2$). 
 To provide a concrete example, at $n=1$, $t^\prime=0$, $T=0$, $U/t=4$, the energy per site for a $20\times 20$ system after ${\Theta}$-averaging is $E/t=-0.86038\pm 3 \times 10^{-5}$. After a weighted least square fitting with $L=4,6, \cdots, 18, 20$, the final result in the thermodynamic limit is $E/t=-0.8603\pm 2 \times 10^{-4}$.

For $n\ne 1$ or $t'\ne 0$, a sign problem appears. The sign problem makes it impossible to converge the ground state projection method for the system sizes and propagation lengths  $\beta$  needed and an alternative method, the   constrained path (CP) approximation is used.  The results reported in this paper follow Ref.~\onlinecite{zhang:2013}, using a trial wave function $|\Psi_T\rangle$ to apply a boundary/gauge condition on the paths that are included in the path integral in auxiliary field space. All results reported here used single-determinant $| \Psi_T \rangle$ with no release. In these calculations,  $|\Psi_T\rangle$ is taken to be a mean-field solution for the Hubbard model for given $U$,  $L$ and ${ \Theta}$ with a $U$ value $U_\text{eff}={\rm min}\{U,4t\}$.  The order parameter in the mean-field solution is chosen to be orthogonal to the spin quantization axis. This choice is found to help preserve symmetry in $| \Psi_T \rangle$, improving the CP results.\cite{shi:2013, shi:2014} The accuracy of the constrained path approach has been extensively benchmarked.\cite{chang:2008, chang:2010, shi:2013, shi:2014} We have carried out additional comparisons with exact diagonalization on $4\times 4$ systems. At $U = 4t$, the relative error on the energy, averaged over 60 randomly chosen $\Theta$ values, is +0.018\% for $n$ = 0.25 and -0.15\% for $n$ = 0.625. At $U/t = 8$ and $n = 0.875$ the relative error is -0.51\% averaged over 20 random $\Theta$ values. We have also verified these estimates in a few systems of larger L, using multi-determinant trial wave functions and constraint release.\cite{shi:2013,shi:2014}

\subsection{Fixed node Diffusion Monte Carlo (FN) with nodes from Variational Monte Carlo (VMC)}

The variational Monte Carlo method constructs a trial wave function that approximates the exact ground 
state of a correlated Hamiltonian.~\cite{gros:1988,gros:1989,yokoyama:1987,yokoyama:1987:2,tocchio:2008}
The wave function depends on parameters that are optimized by minimizing the expectation value of the 
Hamiltonian, which requires a Monte Carlo sampling whenever the trial state is correlated (i.e., it is 
not a simple Slater determinant). We remark that the variational Monte Carlo energy gives an upper bound 
to the exact value and that, with this approach, it is possible to access quite large clusters, with all 
relevant spatial symmetries (translations, rotations, and reflections) preserved. However, it is difficult 
to quantify the systematic errors introduced by the choice of the trial state. Moreover, while spatial 
correlations may be correctly captured, dynamical properties are missed.

We generated simple variational wave functions by applying a density Jastrow factor on top of uncorrelated
states that are built from {\it local} (mean-field) Hamiltonians, containing only few parameters, where the 
physical properties are reflected in a transparent way as different terms inside the variational 
state.~\cite{capello:2005,capello:2006} At finite doping, the uncorrelated states have been obtained from the
BCS Hamiltonian, including superconducting pairing on top of electron hopping; in addition, collinear 
antiferromagnetism with N\'eel order parallel to the $z$ spin quantization axis is also included. At half filling, 
where the system exhibits long-range magnetic order, the uncorrelated state contains only magnetism in the $x-y$ 
plane; in this case, an additional spin Jastrow factor involving the $z$-component of the spin operator is also 
taken. This term couples spins along a direction orthogonal to the magnetic ordering plane, reproducing the 
spin-wave fluctuations above the mean-field state.~\cite{becca:2000} All these variational wave functions with 
Jastrow factors generalize the so-called Gutzwiller state,~\cite{gutzwiller:1963} allowing a description of metals, 
superconductors, and also of Mott insulators.~\cite{capello:2005,capello:2006} Nevertheless, they do not give 
an accurate approximation to the exact ground state in two spatial dimensions, especially close to half filling. 
We obtain a substantial improvement by including the backflow correlations inside the uncorrelated state. On the 
lattice, this corresponds to a redefinition of the single-particle orbitals and it is particularly efficient at 
strong coupling.~\cite{tocchio:2008,tocchio:2011} 

To determine the variational parameters (i.e., the ones related to the Jastrow factors, the ones included in 
the mean-field Hamiltonian, and also those of the backflow correlations), we minimize the expectation 
value of the Hamiltonian. This minimization is performed by constructing a Markov chain using the Metropolis 
algorithm, where walkers are defined by many-body configurations having electrons on lattice sites with given 
spin along the $z$ axis. After performing this optimization, a further improvement can be obtained by applying 
the Green's function Monte Carlo projection technique~\cite{trivedi:1990} to the optimal trial state within a 
fixed-node approximation.~\cite{tenhaaf:1995} 
This procedure allows accurate calculations of the energy and diagonal 
correlation functions, such double occupancies or density-density correlation 
functions. The Ansatz on the nodal structure given by the variational wave 
function induces a systematic error, which cannot be determined a priori 
but can be estimated from the change in energy as the trial wave function 
is improved. We point out that there is a difference between 
continuum and lattice fixed-node approaches. In the continuum, only the 
signs of the trial function are important: if the nodes are correctly placed, 
the exact energy is obtained. By contrast, on the lattice, both the signs 
and the relative magnitudes of the trial function in configurations that are 
connected by a sign flip must be correct in order to have the exact energy.\cite{tenhaaf:1995}
The error bars for finite 
systems are given as the statistical errors of the Green's function Monte Carlo technique and do not include any 
estimates of the systematic errors coming from the fixed-node approximation.

The finite size results are then extrapolated to the infinite system size by using a scaling that depends on the 
carrier concentration. At half filling, we use the $1/N^{3/2}$ scaling (where $N$ is the system size) that is 
appropriate for two-dimensional ordered antiferromagnets.~\cite{neuberger:1989,fisher:1989} In this regime, the 
error bar for the infinite system size is given by a fitting error of the linear regression. At finite dopings, 
the size scaling may suffer from shell effects: a smooth behavior can be obtained only when a sequence of 
closed-shell configurations are taken (i.e., electron fillings for which the non-interacting case corresponds to 
a unique ground state). In the generic case, size effects may be dominated by the ones present at $U=0$. This is 
the case for large dopings (e.g., $n=0.8$) and all interactions, and intermediate dopings (e.g., $n=0.875$) and 
small interactions ($U \lesssim 4$). Here, for every available size, the ratio between the energy at finite $U$ 
and the one at $U=0$ is roughly constant and the thermodynamic limit can be assessed by fitting this ratio,
namely the infinite-size energy is obtained by multiplying the aforementioned ratio by the thermodynamic value at
$U=0$. The extrapolated value is assumed to be normally distributed with an error bar taken as the difference between the estimated thermodynamic limit and the largest available 
size. For intermediate dopings ($n=0.875$), the size scaling starts to deviate from the $U=0$ case around $U/t=4$,
and we decided to take the point at the largest size as the infinite size limit, with an uncertainty of twice the difference to the next lowest system size. We remark that, in this case, a linear regression with the 
$1/N$ scaling gives an estimate of the thermodynamic limit that is compatible (within one error bar) with the 
point at the largest size.

\subsection{Multi-reference Projected Hartree--Fock (MRPHF)}

The multi-reference projected Hartree--Fock method \cite{rodr:2012,rodr:2013,leprevost:2015} is a ground state wave function approach based on a trial wave function $|\Psi \rangle$ characterized by the quantum numbers $\Theta, K$ that is constructed out of a set of broken-symmetry Hartree-Fock wave functions ($\{ |\mathcal{D}_i\rangle \}$) via projection operators. The idea is that a broken-symmetry determinant  includes the dominant correlation physics  while the projection  restores the physical symmetries.  The wave function takes the form
\begin{equation}
  |\Psi^\Theta_K \rangle = \sum_{i=1}^n \sum_{K'} f^{i,\Theta}_{K'}
  \hat{P}^\Theta_{KK'} |\mathcal{D}^i \rangle,
  \label{Eq:mrphf}
\end{equation}
and the parameters $ f^{i,\Theta}_{K'}$ are determined by minimizing
the energy.  The projector $\hat{P}^\Theta_{KK'}$ restores the
symmetries (characterized by the quantum numbers $\Theta,K$) in $|\Psi
\rangle$ and can be formally written as \cite{ring_schuck,schmid:2004}
\begin{equation}
  \hat{P}^\Theta_{KK'} = \frac{h}{L} \sum_m \Gamma^{\Theta\ast}_{KK'}
  (m) \, \hat{R}(m)
\end{equation}
in terms of the rotation operators $\hat{R}(m)$ and the
         irreducible representation matrices $\Gamma^{\Theta} (m)$
         associated with the
elements $m$ of the symmetry group of the problem.  Here, $h$ is the
dimension of the irreducible representation, while $L$ is the volume
of the group. The character of the broken symmetry determinant is
optimized {\em in the presence} of the projection operator ({\em
  i.e.}, a variation-after-projection approach), which results in
broken symmetry determinants with well defined
defects.\cite{rodr:2013}

Our expansion employs Slater
determinants that break the space group and spin ($\hat{S}^2$)
symmetries of the lattice, but preserve $S_z$ symmetry.  All the
broken symmetries are restored using the appropriate projection
operators.  The series of $i$ determinants in Eq. \ref{Eq:mrphf} is
constructed through a chain of variational calculations, using the FED
(FEw-Determinant) approach.\cite{rodr:2012,rodr:2013,schmid:2004} In
this procedure, after a wave function with $n-1$ intrinsic determinants
is available, a wave function with $n$ determinants is variationally
optimized by adjusting the Thouless coefficients determining the last
added determinant. The full set of linear coefficients
$f^{i\Theta}_{K'}$ is re-adjusted.
The MR-PHF approach becomes exact as the number
of determinants retained tends to the size of the Hilbert space,
         {\em i.e.}, as Eq.~(\ref{Eq:mrphf}) becomes a coherent state representation of the
exact ground state wave function.  

If the number of determinants is fixed at a finite, not too large,
value, these calculations can be performed for large lattices with polynomial cost
         ($\mathcal{O}(N)^4$ or so, where $N$ is the number of sites
         in the lattice).
However, the number of determinants required to obtain results with a
given accuracy increases exponentially with increasing system size.  In this
work, implementation aspects have compelled us to keep the number of
determinants roughly constant, so that the quality of the solution
decreases as system size is increased, and, consequently, the energy
increases, precluding a thermodynamic limit extrapolation. 
We have used expansions with 4, 24, and 32 determinants for half-filled
$4\times 4$, $6\times 6$, and $8\times 8$ lattices, respectively. In the
lightly-doped regime, larger expansions have been used: 48 determinants
in a $10\times 4$ ($\langle n \rangle=0.8$) lattice and 80 determinants in a
$16\times 4$ ($\langle n \rangle=0.875$) one.
  A
linear extrapolation in the reciprocal of the number of determinants, i.e. in  $1/n$ has been performed to the infinite
configuration limit for the ground state energies.

The calculations presented could in principle be
improved in a number of ways: additional symmetries could be broken and
restored (such as $\hat{S}_z$ or particle number) in the reference
configurations; more configurations could be included; and/or a full
optimization of all determinants could be performed, in the spirit of
the resonating Hartree--Fock approach.\cite{tomita:2004} The accuracy
of any one result can therefore be increased, as shown by
Rodr\'iguez-Guzm\'an {\em et al}. \cite{rodr:2014} or by Mizusaki and
Imada \cite{mizusaki:2004} in the closely related path integral
renormalization group (PIRG) approach. Recently, Tahara and Imada
\cite{tahara:2008} have combined the symmetry-projected determinant expansion 
with short- and long-range Jastrow factors within a
variational Monte Carlo framework, which may be used to further
increase the accuracy.  These techniques, and others, have been used to explore, for example, spin and charge stripe phases, which we do not explore in this work.\cite{carlos:2015,corboz:2014,corboz:2011}

\subsection{Unrestricted Coupled Cluster - Singles and Doubles (UCCSD)}

Coupled cluster (CC) theory\cite{Paldus1999,Bartlett2007,ShavittBartlett} is a ground-state wave function technique.  It is widely used in quantum chemistry and often considered the best source of precise data for molecules that are neither too large nor too strongly correlated. 
Its application to the Hubbard model has been more limited, where it has been used in two 
different formulations. 
In the first form, which directly exploits the translational invariance of the 
lattice to work in the thermodynamic limit, the theory is formulated in the site-basis, 
starting from an infinite N\'{e}el ordered reference, from which clusters of excitations
which change occupancy and flip spins are created \cite{petit1994coupled,BishopHubbard}. In the second form, the theory starts from a single determinant reference state 
on a finite lattice, from which clusters of particle hole excitations are created \cite{asai1999coupled}. 
This is similar to how the theory is used in quantum chemistry, and is the 
formulation discussed further here. As this second form does not 
work in the thermodynamic limit, the energies must be extrapolated.

The CC wave function is written as $|\Psi\rangle = \exp(\mathrm{T}) \, |0\rangle$ where
$|0\rangle$ is  a single determinant reference state, and  $T = \sum_n T_n$, where $T_n = \sum_{q_n} t_{q_n} \, A_{q_n}^\dagger$ is the cluster operator.  The operator $A_{q_n}^\dagger$ creates an excited determinant $|\Phi_{q_n}\rangle$ which contains $n$ particle-hole pairs relative to the reference state. In its standard and simplest version (used here), the energy and coefficients $t_{q_n}$ are obtained by solving the Schr\"odinger equation projectively:
\begin{subequations}
\begin{align}
E &= \langle 0 | \mathrm{e}^{-T} \, H \, \mathrm{e}^T |0\rangle,
\\
0 &= \langle \Phi_{q_n} | \mathrm{e}^{-T} \, H \, \mathrm{e}^T |0\rangle \qquad \forall q_n.
\end{align}
\end{subequations}
CC theory thus diagonalizes a similarity-transformed Hamiltonian $\bar{H} = \exp(-T) \, H \, \exp(T)$ in a subspace of states defined by $|0\rangle$ and $|\Phi_{q_n}\rangle$.  Note that because $T$ is a pure excitation operator, the commutator expansion used to evaluate $\bar{H}$ truncates after four commutators.

If the sum over $n$ in defining the cluster operator is carried out to all orders, the exact ground state wave function is reproduced.  In practice the operator $T$ is typically restricted to terms involving a small number of particle-hole pair excitations above the reference state ($n \leq n_{\text{max}}$) and $|\Psi\rangle = \exp(\mathrm{T}) \, |0\rangle$ is projected onto the space of states with up to $n_{\text{max}}$ particle-hole pairs; the accuracy then depends upon $n_{\text{max}}$ and the choice of reference state $|0\rangle$.  In a lattice model such as the Hubbard model with $N$ sites, the computational cost is, roughly speaking, proportional to $N^{2 (n_{\text{max}}+1)}$. 

The calculations reported in this manuscript primarily limit $T$ to $n\leq 2$, i.e. to the creation of  only singly- and doubly-excited determinants, giving what we refer to as CC with single and double excitations (CCSD).\cite{CCSD,CCSDGES}  For select examples, we have included corrections for triple and occasionally quadruple excitation effects. The accuracy of CC theory, and the need for higher excitations, depends on how well the reference $|0\rangle$ captures the qualitative physics. When the reference is accurate, then singles and doubles excitations may be sufficient; 
however,  when the reference determinant bears little resemblance to the exact wave function (as may happen in strongly correlated systems) a much higher degree of excitation is required to recover the correct physics. For this reason, the calculations reported here use a symmetry-broken unrestricted Hartree-Fock (UHF) reference determinant, because UHF can provide a better mean-field description, particularly near half filling where antiferromagnetic correlations dominate; this defines the unrestricted CCSD (UCCSD) method used here.  Note, however, that particularly for doped systems with large $U$ there are a plethora of nearly degenerate UHF states, and finding the best reference for UCCSD is not straightforward.  We have prepared the UHF solution following the prescription of Ref.~\onlinecite{ShiweiUHF}.  In principle the deficiencies of the reference determinant can be corrected in what is known as Brueckner CC\cite{Handy1989} where the reference determinant is adjusted to eliminate single excitation effects. These calculations are more computationally demanding and we have not pursued them here.

An important virtue of the exponential parametrization of the wave function is that the CC energy has
 a non-trivial thermodynamic limit even for restricted excitations $n_{max}$~\cite{Paldus1999,Bartlett2007,ShavittBartlett}  Thus, as lattice size increases, the energy per site approaches the thermodynamic limit for the given $n_{max}$, and the exact thermodynamic limit as $n_{max}$ is increased.  For smaller values of $U$ where convergence to the thermodynamic limit is slower, we have converged second-order perturbation theory out to the thermodynamic limit and added a correction for the difference between CC theory and perturbation theory which, for small $U$, converges quickly with respect to system size. Double occupancies have been computed by numerical differentiation of the CC energy with respect to $U$.
We also provide UCCSDTQ estimates of the ground state energy (labelled as
UCCSDTQ*) for n=1 systems, where the triples (T) correction is obtained
as UCCSDT-UCCSD energies for a $6\times6$ system, and the quadruples (Q)
correction is obtained from UCCSDTQ-UCCSDT energies for a $4\times4$ lattice.
For n=0.8 (n=0.875), our UCCSDT estimates (labeled as UCCSDT*) are
obtained from $10\times4$ ($16\times4$) lattices. 

The UCCSD calculations reported here can be completed in a few hours using standard quantum chemistry packages.\cite{GDV}  Even at half filling where there are not many Hartree-Fock solutions to be concerned with, we find large effects due to single excitations, which suggests that the coupled cluster calculations could be substantially improved by optimizing the identity of the reference determinant.  Similarly, we generally see significant corrections due to triple and higher excitations; these can also be computed with standard packages,\cite{Kallay} but while optimizing the reference determinant and including higher excitation effects will increase the accuracy of the coupled cluster calculations, they also increase the cost.  

\subsection{Density Matrix Renormalization Group (DMRG)}\label{sec:DMRG}
The density matrix renormalization group\cite{white:1992} is a variational ground state wave function technique.  It constructs the ground state of a system by diagonalizing the Hamiltonian in a finite subspace spanned by  an iteratively constructed basis that is optimized via a Schmidt decomposition that minimizes the spatial extent of the quantum mechanical entanglement between basis states. 

DMRG is generally believed to be the optimal method for finding grounds states of one dimensional lattice models.  In the application to one dimensional systems two sources of error must in principle be controlled. Results for a fixed system length of $L$ must be converged with respect to basis size $m$, and then the converged results must be extrapolated to $L\rightarrow\infty$. However, for most of the one dimensional Hamiltonians of current interest the convergence is very rapid and in practice large enough $m$ and $L$ are accessible numerically, so extrapolation is not required.   

Application of DMRG to a finite size 2D system proceeds by defining an effective one dimensional problem, to which the standard one-dimensional DMRG is applied.  Two dimensional tensor network generalizations of the DMRG ideas have attracted tremendous recent interest, but these methods have not yet produced results for the two dimensional Hubbard model that can be included in the present comparison. \cite{verstraete:2004,vidal:2007}

Most current implementations of DMRG require  open boundary conditions. The canonical method for creating an effective one dimensional system from a finite-size  two dimensional one is to impose periodic boundary conditions in one direction and open boundary conditions in the other, thereby defining a cylinder of finite length and finite circumference. One then defines an effective one dimensional problem by  indexing the sites along a one dimensional path that covers all of the sites on the cylinder and taking the matrix elements of the Hamiltonian in this basis.  The price is that two sites separated by a small distance along the cylinder axis in the physical system are separated by a distance of order the cylinder circumference in the effective one dimensional model. The effective one-dimensional problem thus has long ranged terms, which imply longer ranged entanglement and require that more states are kept in the optimal basis. The number of states needed grows exponentially with the circumference of the cylinder (width of original finite lattice) meaning that there is a sharp cut-off in the accessible system widths, typically around width 6 in the Hubbard systems; however, systems of very large cylinder length can be studied.  The extrapolation to the thermodynamic limit must thus be handled with care.

The DMRG calculations reported here were performed with the standard DMRG finite system algorithm\cite{white:1992,white:1993}  Two types of cylinders were considered: one with the axial and circumferential directions aligned parallel to the bonds of the square lattice and one rotated by 45 degrees.  When one cuts a cylinder in two, the 45 degree rotated system has fewer sites on the boundary per unit length, and thus one expects a smaller growth of the entanglement with the length of the cut (governed by the area law).  For the undoped antiferromagnetic system, the rotated system also is unfrustrated both for odd and even circumferences, reducing shell effects in the finite size results.  For half filled systems both types of orientation are considered and they show good agreement, and error bars are estimated to incorporate both the error bars on the data points for specific widths, and the differences between the two orientations. However we note that for the half-filled systems, better results were obtained with the rotated system.\cite{white2007neel,stoudenmire:2012}  With doped systems, we  see striping behavior at stronger coupling.  A stripe is a line of holes which act as a domain wall in the antiferromagnetism on either side. These stripes have been seen in the t-J model with DMRG starting in the late 90's.\cite{white:1998} The stripes typically wrap around the cylinder, with a specific even number of holes in the ring-stripe. With doping, the optimal number of holes can change.  Striped configurations with the wrong number of holes in a stripe are typically metastable. For the ordinary orientation, we were able to sort out stripe fillings, finding the low energy states and avoiding metastability.  For the 45 degree rotated lattices, the patterns seem more delicate, and we have not yet sorted out the lowest energy configurations. We thus present only the results for standard orientation for doped systems. 

Convergence issues pose more severe problems than in the standard one dimensional cases, both because of the intrinsic limits on system size discussed above and because in some cases the presence of several metastable states can cause troubles for extrapolation, and can lead to the appearance of states that are not the ground state and may be important only for finite systems. In the systems studied here, metastability was traced to the presence of physically different ``striped'' states for different hole doping in the DMRG cylinder.
 
We obtained converged results as follows. For each cylinder of a specific length $L$, we extrapolated the energy in the number of basis states, $m$.  To make this extrapolation reliable $m$ was slowly increased, but each $m$ was used for two consecutive sweeps.  The truncation error and the energy were
measured on the second sweep for each $m$. Then a linear extrapolation of energy versus truncation error is used to obtain the ground state energy
with error bars.
The deterministic nature of DMRG can result in uncertainties due to fitting which do not appropriately represent the uncertainty in choice of extrapolation procedure.  We therefore assume a normally distributed error of 1/5 the difference between the last point and the extrapolated value which we justify by comparison to the accuracy of previous DMRG data.\cite{white2007neel,stoudenmire:2012} Metastability was signaled by lack of   linearity in this extrapolation. When this was found we determined which  state had the lowest energy.  The system was then rerun with an initial state favoring the lower energy state producing results in the lower energy configuration such that the extrapolations are linear in the truncation error.

For a fixed width we then extrapolate the energies linearly in  $1/L$ to get an energy per site for an infinite cylinder.  Errors are estimated statistically using the error bars on each point. To reduce the finite size
effects from different widths, we employ a simple version of the phase averaging trick\cite{lin:2001} by taking an average over  periodic and antiperiodic boundary conditions around the cylinder. The phase averaging eliminated an oscillation in the energy as a function of system width in the 45 degree rotated systems, and in general nicely accelerated convergence. We then analyzed the results as a function of cylinder width. For half filling we found that extrapolating the energy to the thermodynamic limit by $1/{\mathrm{width}^3}$ works well.  This is the finite-size behavior in the Heisenberg model, where it is well understood.\cite{Sandvik97}

\subsection{Density Matrix Embedding Theory (DMET)}

The density matrix embedding theory (DMET)~\cite{Knizia2012,Knizia2013} is a ground state embedding method formulated in terms of wave function entanglement.  Given an impurity cluster of size $N_c$,  DMET maps an $L\times L$ bulk many-body problem ($L$ is chosen very large) to an  impurity and bath many-body problem, yielding a problem with $2N_c$ sites in total.
The mapping is constructed from the Schmidt decomposition~\cite{Peschel2012} of a bulk wave function 
$|\Psi\rangle$. The formulation is exact if $|\Psi\rangle$ is the exact bulk ground state, or in the limit of  impurity cluster size $N_c\to \infty$.
Since the exact bulk state is not known a priori, an approximate bulk state is used 
for the impurity mapping. 
Recently, DMET has been applied to both ground-state and linear response spectral properties of the Hubbard model\cite{Knizia2012,Booth2013,Bulik2014,Chen2014}.
In this work we use a general BCS bulk state, the ground-state
of the DMET lattice Hamiltonian given by the hopping part of the Hubbard Hamiltonian
augmented with the DMET correlation potential $u$, that is applied in each cluster supercell of $N_c$ sites in the bulk lattice. This bulk state is allowed to spontaneously break spin and particle number
symmetry through the self-consistency cycle that determines $u$ (which
contains both particle number conserving and non-conserving terms). In this cycle, 
the bulk state $|\Psi\rangle$ is updated from the interacting
impurity and bath solution $|{\Phi}\rangle$,
by minimizing the difference between $|\Psi\rangle$ and $|\Phi\rangle$ (as measured by their (generalized) one-particle density matrices) with respect to the potential $u$.

For the bulk lattice, we use $L=72$. From calculations on larger $L$, we find that the finite lattice error associated with this choice is negligible, on the scale of the significant digits reported. 
 The BCS bulk state is obtained
by solving the lattice
spin-unrestricted Bogoliubov-deGennes equation~\cite{Gennes1966,Yamaki2004} with the correlation potential $u$.
The impurity and bath problem is solved in the BCS quasiparticle basis, with general one-body and two-body interactions that do not conserve particle number or locality.
We use the density matrix renormalization group (DMRG)~\cite{white:1992}
as an impurity solver (adapted from the quantum chemistry DMRG code \textsc{BLOCK}~\cite{chan2002highly,chan2004algorithm,Chan2011}), with a maximum number of renormalized
states $m$=2000 (DMET self-consistency is performed up to $m$=1200). 
The DMET lattice  and impurity Hamiltonians are augmented with a chemical potential $\mu$, to ensure that the relative error in 
particle number is less than $0.05\%$.
We used impurity clusters of dimensions $2\times2$, $4\times2$, $4\times4$ and $8\times2$ at each point in the phase diagram.
The energies and observables were then extrapolated to the thermodynamic limit 
using a linear relationship with $N_c^{-1/2}$, as appropriate to a non-translationally invariant cluster embedding theory.

The total DMET uncertainty is estimated by combining the errors from three sources (i)  convergence of DMET self-consistency, 
(ii)  solution of the impurity many-body problem using DMRG, (iii)  extrapolation to the limit of infinite impurity size. 
We estimate the self-consistency error (i) using the difference of the last two DMET self-consistent iterations. The average 
self-consistency error is below $5\times10^{-4} t$ in the energy for all cluster sizes. The impurity solver error (ii) is from using a finite number of renormalized
states $m$ in DMRG. This error is only non-zero in  clusters larger than $2\times2$. The energy and observables are extrapolated to $m=\infty$ 
using the standard linear relation  between energy and DMRG truncation error.\cite{legeza1996accuracy,white2007neel,chan2002highly} For $4\times4$ impurity clusters, the truncation error is large enough to
contribute also to the converged DMET self-consistent correlation potential $u(m)$. To take this into account, we extrapolate using (a) self-consistent DMET results converged at different $m$, and
(b) non-self-consistent DMET results using different $m$ in the DMRG impurity solver at a fixed correlation potential $u(m_{\max} )$,
where $m_{\max}$ is the maximum $m$ used in the DMET self-consistency.
The difference between the two extrapolations is then added to the total DMET uncertainty.
For the cluster sizes in this study the errors due to (i) and (ii) are small and easily controlled.  Therefore, the finite size impurity error is the main source of uncertainty at almost all points in the phase diagram.  It is estimated as the standard deviation of the finite size extrapolation.
The quality of the approximate DMET impurity mapping depends on the approximate lattice wave function $|\Psi\rangle$ and
decreases as the coupling strength $U$ increases, especially for carrier concentrations near half filling. This 
slows down the cluster size convergence of the results for large $U$, increasing the uncertainty.
In the strong coupling, weakly doped region, we find competing homogeneous and inhomogeneous orders that become
very sensitive to the cluster size and shapes (similar to the stripes
observed in DMRG). It is difficult to 
reliably extrapolate these results to the thermodynamic limit.
As a result, the total DMET uncertainties range from about $10^{-4}t$ at half filling and low densities to a maximum of about $10^{-2}t$ in the strong coupling, underdoped region.
A detailed description of the methodology and extrapolation procedures for the calculations is contained in Ref.~\onlinecite{zhengchan}.

\subsection{Dynamical Cluster Approximation (DCA)}
The dynamical cluster approximation (DCA)\cite{hettler:1998, hettler:2000,maier:2005} is an embedding technique in which an approximation to the electron self energy is obtained from the solution of a quantum impurity model consisting of some number $N_c$ of interacting sites coupled to an infinite bath of non-interacting electrons. In applications to the Hubbard model, the interactions in the impurity model are the interactions of the original problem while the one-body terms are determined from a self-consistency condition relating the Greens functions of the impurity model to those of the lattice.    The DCA is a particular generalization to $N_c>1$  of the `single site' dynamical mean field method.\cite{georges:1992, georges:1996} Other generalizations have also been introduced,\cite{lichtenstein:2000,kotliar:2001} but results from these methods are not considered here. The single-site method was  motivated by the observation that in an appropriately defined infinite coordination number limit  an exact solution of the Hubbard model can be found,\cite{metzner:1989} and can be recast in terms of the solution of a single-site impurity model.\cite{georges:1992} It was later understood that generalization to $N_c>1$  impurity model representations allows a treatment of the finite coordination number model that becomes exact as the number of impurity sites $N_c\rightarrow \infty$.

The DCA formulation partitions the Brillouin zone into $N_c$ equal area tiles and approximates the self energy $\Sigma$ as a piecewise constant function of momentum taking a different value in each tile:
\begin{equation}
\Sigma(\mathbf{k},\omega)=\sum_{a=1...N_c}\phi_a(\mathbf{k})\Sigma_a(\omega)
\label{DCAeq}
\end{equation}
with $\phi_a(\mathbf{k})=1$ for $\mathbf{k}\in a$ and $\phi_a(\mathbf{k})=0$ for $\mathbf{k}\notin a$. The tiles, $a$, map directly on to impurity model single-particle levels, the impurity model Greens function and self energy are diagonal in the impurity model $a$-basis and the DCA self-consistency equation 
\begin{equation}
G^{imp}_a(\omega)=\int_a\frac{d^2k}{(2\pi)^2}\frac{1}{\omega-\varepsilon_k-\Sigma_a(\omega)}
\label{DCASCE}
\end{equation}
determines the remaining parameters of the impurity model. The self consistency loop is solved by iteration; an initial guess for the impurity model parameters produces a set of $\Sigma_a$ which are then used to update the impurity model parameters. The loop typically converges well and errors associated with the self consistency are smaller than errors in the solution of the impurity model. 

We obtain results for different $N_c$ in the paramagnetic phase and extrapolate to the thermodynamic limit by exploiting the known\cite{maier:2005} $N_c^{-d/2}$ scaling for momentum averaged quantities in $d$-dimensions and systematically increasing $N_c$.\cite{fuchs:2011, leblanc:2013}

To solve the $N_c$ site impurity problem we use a CT-AUX algorithm\cite{gull:2008,gull:2011} with submatrix updates.\cite{gull:2011:smu} Our codes are based on the ALPS libraries.\cite{ALPS20,ALPS_DMFT} Selected points have been compared to a CT-INT\cite{rubtsov:2005} implementation based on the TRIQS libraries.\cite{parcollet:2015}

In this work we provide extrapolated DCA results from clusters of sizes $N_c$=$16$, $20$, $32$, $34$, $50$, $64$, $72$ and $98$, depending on temperature and densities, in order to give a reliable estimate of the properties of the 2D Hubbard model in the thermodynamic limit. 

The CT-AUX method is a type of diagrammatic Monte Carlo. In the $T>0$, impurity model context the diagrammatic series is absolutely convergent and the issues discussed below for Diag-MC in the infinite-lattice context are not important. However,  the CT-AUX method is restricted at low $T$ by the existence of a Monte Carlo sign problem in the auxiliary field solver. The sign problem  worsens rapidly as $U$ is increased, as $T$ is decreased, or as $N_c$ is increased and  is particularly evident in the density range $n=[0.8, 1.0)$.  Further, as temperature is decreased, the length scale of correlations in the system increases, resulting in larger finite size effects. We take a conservative approach to determining the uncertainty in the extrapolation procedure.  We include both a statistical uncertainty in the extrapolation in $1/N_c$ as well as an additional uncertainty which we take as half  the difference of the extrapolated value from the largest $N_c$ value explored.  This gives a robust representation of an extrapolation error which is larger when finite size effects are large, but that also vanishes as $N_c$ increases. As such, the error bars for extrapolation of our DCA results to the thermodynamic limit  contain both stochastic and finite size uncertainties, and values for finite system sizes with stochastic error bars are provided.

\subsection{Dual Fermion (DF) Ladder Approximation}
The dual fermion approach\cite{Rubtsov2008,Rubtsov2009,Hafermann2012} is a diagrammatic extension of the single site dynamical mean field theory (DMFT). The DF technique is motivated by the idea that non-local corrections to DMFT can be captured by a perturbative expansion around a solution of the dynamical mean field equations. In formal terms the expansion requires summing a series of diagrams for two and higher particle correlations, with vertices defined in terms of the fully interacting but {\em reducible} vertices of the impurity model.
In this regard, the DF technique is similar in spirit to methods such as the dynamical vertex approximation ($D\Gamma A$) and dynamical mean-field extensions of fRG (DMF$^2$RG) which approximate interactions on the level of 2-particle vertex functions.\cite{toschi:2007,taranto:2014,wentzell:2015}
 In practical implementations to date  the dual fermion expansion is truncated at the two particle level (higher than two-particle interactions are omitted) and the series of two particle corrections is approximated by a few low-order terms or a ladder resummation. One of its strengths lies in the ability to describe phase transitions of the system by employing resummations of the relevant diagrams.\cite{Antipov2014b,Otsuki2014}

The DF results presented here are obtained using the open-source \texttt{opendf} code \cite{OpenDFZenodo}, starting from a single-site ($N_c=1$)  dynamical mean field solution obtained with a continuous-time auxiliary-field (CT-AUX)\cite{gull:2008,gull:2011,gull:2011:smu} impurity solver. The method is limited by the accuracy to which reducible impurity vertex functions can be obtained, which is a polynomial (cubic) complexity problem.  Within the approximation of neglecting higher order vertices and only considering ladder diagrams, a systematic estimation of deviation from the true interacting system is not possible, and we omit error bars altogether.

\subsection{Diagrammatic Monte Carlo (DiagMC)}
The diagrammatic Monte Carlo method (DiagMC) begins from the observation that within standard many-body perturbation theory, any quantity $Q$ that depends on some set of arguments ${\mathbf y}$ (which may include both continuous components such as frequency and momentum  and discrete components such as spin) may be  expressed as an infinite  series of terms, each of which involves multi-dimensional integrals and sums over many variables:
\begin{equation}
Q({\mathbf y}) = \sum_{\alpha=0}^{\infty} \sum_{\xi } \int \! \dots \! \int
d{\mathbf x}_1 \! \dots d{\mathbf x}_{\alpha}
D(\alpha,\xi , {\mathbf x}_1, \dots , {\mathbf x}_{\alpha} ; {\mathbf y} ) \;,
\label{eqn:diagmc}
\end{equation}
Here the $D$ are known functions defined by the Feynman rules of diagrammatic perturbation theory. The series order $\alpha$ controls the number of internal integration/summation variables $ \{ {\mathbf x}_1, \dots , {\mathbf x}_{\alpha} \}$, and $\xi $ labels different terms of the same order in the series. The quantity $Q$ may be  the electron Greens function $G$, the self-energy, $\Sigma$, the screened interaction, $W$, the polarization operator, $\Pi$, the pair-propagator (for contact interaction), $\Gamma$, and its self-energy, $\Sigma_{\Gamma}$, or the current-current or other correlation functions. Basic thermodynamic properties (particle, kinetic, and potential energy densities)  in the Grand Canonical ensemble are readily obtained from  $G$ and $\Sigma$, see Ref.~\onlinecite{FW}.

The most widely-used formulation of perturbation theory is in terms of Feynman diagrams. In this case standard rules relate the graphical representation of a given term in the series to the corresponding mathematical expressions for the $D$, which are typically given (up to a sign or phase factor)  by a product of functions associated with  graph lines, $D=\prod_{\rm lines} F_{\rm line} ({\mathbf x}_{\rm line})$. In perturbation theory for particles interacting via pairwise forces, these lines are associated with the interaction potential, $U$, and bare single-particle propagators $G^{(0)}$.  If we denote the collection of all external and internal variables
that allow for a complete characterization of the diagram (diagram topology and internal and external variables) as $\nu \equiv (\alpha,\xi , {\mathbf x}_1, \dots , {\mathbf x}_{\alpha} ; {\mathbf y})$, then Eq.~(\ref{eqn:diagmc}) can be viewed as weighted average over the configuration space $ \{ \nu \}$: i.e., $Q=\sum_{\nu} D_{\nu} \equiv \sum_{\nu} e^{i\varphi_{\nu}} |D_{\nu}|$, where the modulus of $D_{\nu}$ defines the configuration 'weight', and $\varphi_{\nu} = \arg D_{\nu}$. The basic idea of the diagrammatic Monte Carlo (DiagMC) method is to use stochastic sampling of the configuration space to compute $Q$ by treating $\propto |D_{\nu}|$ as the 
probability density for the contribution of  point $\nu$ to the sum. In DiagMC the diagram order, its topology, and all internal and external variables are treated on equal footing and each diagram represents a point, not an integral, in $\{ \nu \}$. The MC process of generating diagrams  with probabilities proportional to their weight is based on the conventional Markov-chain updating scheme \cite{prokofiev:1998, Mishchenko:2000, vanhoucke:2010} implemented directly in the space of continuous variables. 

Since connected Feynman diagrams are formulated directly in the asymptotic limit, there is no infinite system size limit to take. The main numerical issue concerns the convergence of the Monte Carlo process, which is complicated by the exponential proliferation of number of diagrams with perturbation order $\alpha$. This leads to exponential computational complexity since final results are recovered only after extrapolation to the infinite diagram-order limit.   The fermion sign enters the calculation in an interesting way: different diagrams  have different signs (arising from the different orderings of fermion creation and annihilation operators) and at large diagram order the contributions of diagrams with plus and minus sign tend to cancel. This cancellation is in fact responsible for the convergence of the many-body perturbation theory.\cite{vanhoucke:2010,prokofiev:2007} To manage this issue it is useful to consider a Monte Carlo process for approximate series in which the maximum diagram order is limited to some finite value via a hard or soft cutoff, and then cutoff is increased until convergence is reached.

Diagrammatic Monte Carlo techniques can also take  advantage of known field-theoretical techniques to  run the calculation in a self-consistent mode in which certain infinite  series of diagrams are summed and then  automatically absorbed into the renormalized propagators and interaction lines using Dyson-type equations. One example is the skeleton expansion  \cite{prokofiev:2007}; another is a ``bold'' expansion in perturbative corrections to an analytic partial resummation.\cite{vanhoucke:2012,gull:2010:bold} This flexibility allows for an additional control over systematic errors coming from series extrapolation as well as convergence issues---different schemes should produce consistent final results. 

In this work we employ four complementary techniques:

$\bullet$ A $[G^{(0)}]^2U$-scheme  based on a Taylor series expansion of $\Sigma$ in powers of $U$ with fixed
shifted chemical potential $\tilde{\mu}=\mu-Un/2$, see Refs.~\onlinecite{vanhoucke:2010,kozik:2010}. The total electron density, $n=\sum_{\sigma} n_{\sigma}$, and the chemical potential are computed {\it posteriori} (after results are extrapolated to the $\alpha_{max} \to \infty$ limit).

$\bullet$ A $G^2W$-scheme based on skeleton series for $\Sigma$ and $\Pi$ in which all lines in the diagram are understood as fully dressed Green's functions and screened interactions. Self-consistency is implemented by feedback loops when $G$ and $W$ are obtained by solving algebraic Dyson equations, $G^{-1}=[G^{(0)}]^{-1}-\Sigma$ and $W^{-1}=U^{-1}-\Pi$, in momentum-frequency representation, see Refs.~\onlinecite{kulagin:2013, kulagin:2013:prb, mishchenko:2014} for more details.

$\bullet$ A $G^2\Gamma$-scheme  based on the skeleton series for $\Sigma$ and $\Sigma_\Gamma$
when all lines in the graph are understood as fully dressed single-particle (Green's functions) and two-particle propagators. This compact formulation is possible only for a contact interaction potential when the sum of ladder-type diagrams for spin-up and spin-down particles has the same functional structure as the single-particle propagator, see Refs.~\onlinecite{vanhoucke:2012, vanhoucke:2013, deng:2014}. Again, self-consistency is implemented by feedback loops using Dyson equations, $G^{-1}=[G^{(0)}]^{-1}-\Sigma$ and $\Gamma^{-1}=[\Gamma^{(0)}]^{-1}-\Sigma_{\Gamma }$, where $\Gamma^{(0)}$ is the sum of bare ladder diagrams.

$\bullet$ A $[G^{(0)}]^2 \Gamma^{(0)}$-scheme  based on diagrams expressed in terms of bare single- and pair-propagators with shifted chemical potential $\tilde{\mu}=\mu-Un/2$; this is similar in spirit to the $[G^{(0)}]^2 U$-expansion, but with one extra geometrical series (bare ladder diagrams) being accounted for analytically.


To establish the parameter region where DiagMC works, we performed calculations using all four schemes. 
The results were compared to each other and to those obtained by DCA.  Additional insight was also gained by doing calculations in the atomic limit.\cite{kozik:2015} We find that for bare coupling $U/t < 4$ and temperature $T/t >0.1$, all schemes  produce consistent results within statistical errors. At half filling, $n=1$, and $U/t \leq 6$,  the $(G^{(0)})^2U$- and $(G^{(0)})^2 \Gamma^{(0)}$-schemes still produce results consistent with those obtained by DCA and the determinant Monte Carlo method. In the dilute region (small filling factors), the $G^2\Gamma$- and $(G^{(0)})^2 \Gamma^{(0)}$-schemes can be applied for larger values of $U$. 
For dilute systems our benchmarks include points $(U/t=6,n=0.6)$, $(U/t=6,n=0.3)$, and $(U/t=8,n=0.3)$. 

\section{Extrapolations and Uncertainties}
\subsection{Extrapolations}\label{sec:extrapol}
All of the numerical methods we have considered rely on the extrapolation of results to a thermodynamic or asymptotic limit. For DiagMC, which is formulated directly on an infinite system, the extrapolation is in diagrammatic order. All other methods are extrapolated from a finite embedded system, finite cluster, or cylinder with finite width to the infinite system size limit. In many cases, a considerable contribution to our errors comes from this extrapolation procedure, which differs from method to method.
In some cases determining stochastic uncertainties in extrapolation are not possible, in which case we produce estimates of uncertainties by choosing a reference system for a given technique.  We then assume a normal distribution of uncertainty with respect to the reference.
 Specifics of the extrapolation procedure for each system (and of the associated procedure for estimating extrapolation uncertainties) are described in  Sec.~\ref{methods}. 
All methods have therefore defined procedures to estimate error in TL quantities as accurately as possible through the use of known reference systems.  Additional uncertainties due to extrapolation, curve fitting, truncation in excitation order are addressed on a per-technique basis. These added uncertainties are assumed to be normally distributed and defined such that they can, in principle, be made arbitrarily small by adding additional data.
This section illustrates these extrapolations and presents some of the challenges  encountered in performing them.  

\begin{figure}[tbp]
\includegraphics[width=\linewidth]{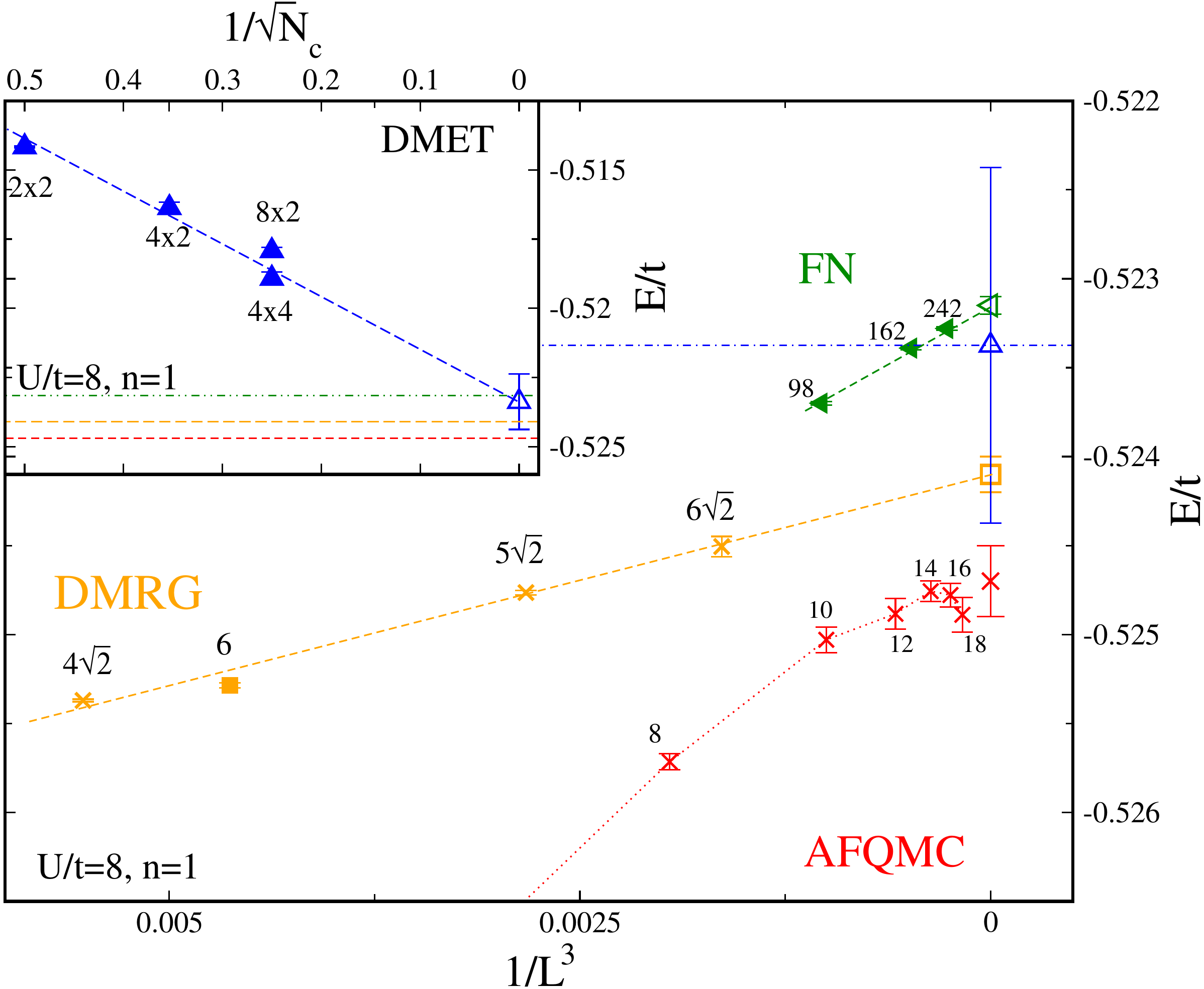}
\caption{\label{fig:extrap_U8n1} (Color online) Extrapolations of the ground state energy at $U/t=8$, $n=1$. Main panel:  AFQMC and FN extrapolated as a function of the inverse cube of the system's linear dimension, $L$, along with DMRG extrapolated in cube of inverse cylinder circumference (also denoted $L$).  DMRG data are presented both for rotated (with $\sqrt{2}$) and unrotated wrapping of cylinders. Inset: DMET data for clusters of size and geometry indicated, plotted against the reciprocal of the  square root of the total number of sites in the cluster $N_c$.}
\end{figure}

\begin{figure}[tbp]
\includegraphics[width=\linewidth]{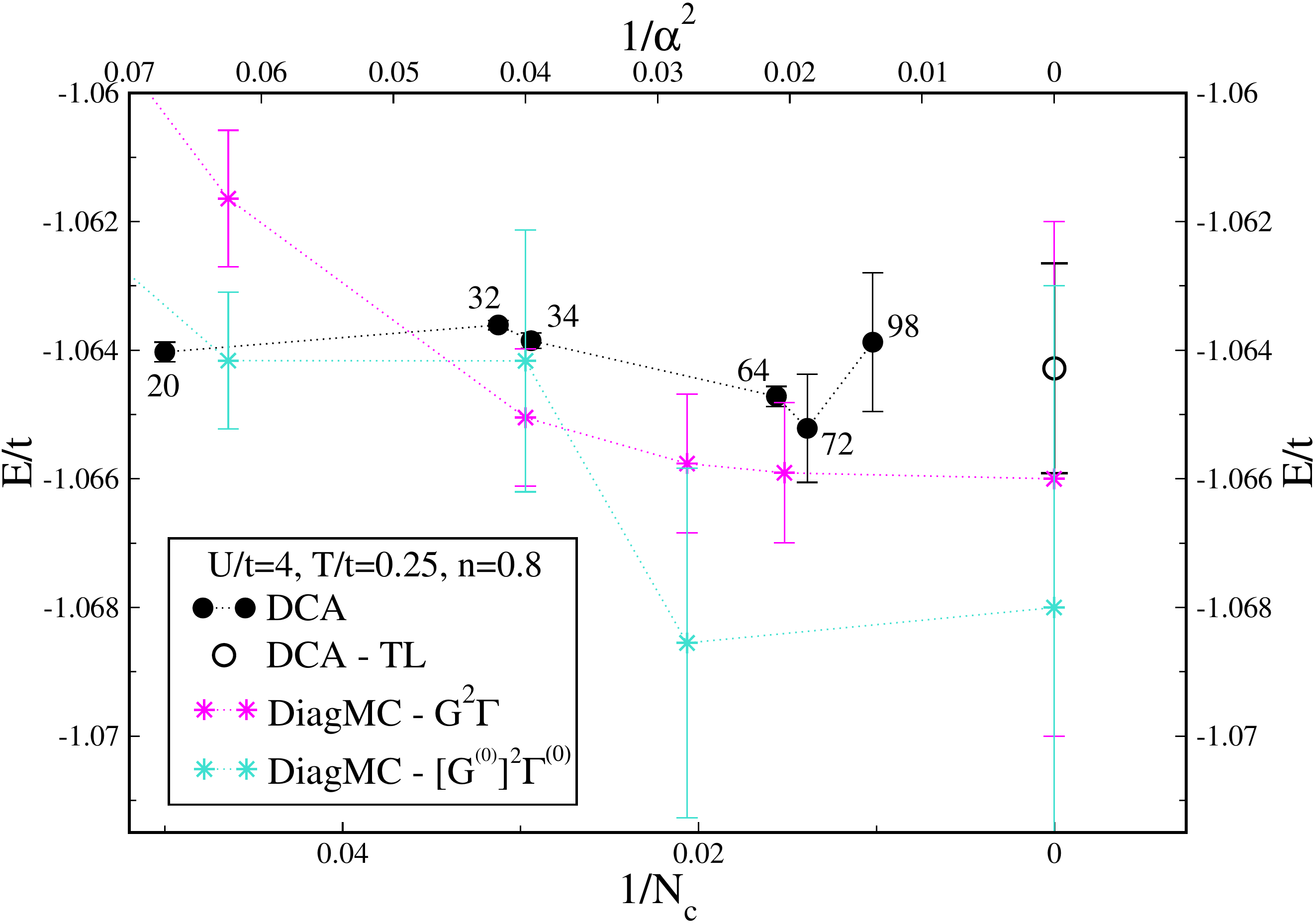}
\caption{\label{fig:extrap_DCA_DIAG} (Color online) Energies obtained at $U/t=4$, $T/t=0.25$ and $n=0.8$ from  $G^2\Gamma$ (magenta) and $[G^{(0)}]^2\Gamma^{(0)}$ (turquoise) as a function of the inverse square of the diagram order parameter  $\alpha$ (upper axis label) along with  DCA results obtained from finite clusters plotted against the inverse of the cluster size    $N_c$ (lower axis labels).}
\end{figure}


We start our discussion with ground state properties. In Fig.~\ref{fig:extrap_U8n1},
DMRG, FN and AFQMC results are presented in the main panel. The system size dependence in all three techniques is clearly visible and  the difference between the estimated thermodynamic limit value and the largest size considered is in some cases outside of the error bars of the thermodynamic limit value: in other words, extrapolation is essential for obtaining the thermodynamic limit value. For this reason, in the results sections we typically present both the thermodynamic limit value and the sequence of finite size results which led to it, so that the reader can see how large an extrapolation is required.  

For DMRG, results from both un-rotated cylinders (filled symbol; other smaller-L data are not shown but lie on the same scaling curve) and rotated cylinders (crosses) are consistent, both scaling as $1/L^3$, although with different slopes, allowing for a clean extrapolation to the thermodynamic limit.
On the other hand,  the AFQMC data indicate a change in scaling for system sizes larger than  $10 \times 10$ geometry under twist averaged boundry conditions. This could indicate either  that unidentified complications occur  in the large system  AFQMC calculations or that  deviations from the $1/L^3$ size dependence might occur in the DMRG data at larger cylinder size (i.e. that the DMRG error bar is underestimated).  In this regard, it is important to note that  the ground state energy of the largest system examined in DMRG, rotated $6 \times \infty$, is within uncertainty of the extrapolated AFQMC data.

The FN data also demonstrate a systematic dependence of the energy on system size,  allowing a precise thermodynamic limit extrapolation.  The deviation of the FN results  from the AFQMC and DMRG results is caused by a systematic fixed node error, which by comparison to other methods seems to be no more than $2\times 10^{-3} t$.

Shown in the inset of Fig.~\ref{fig:extrap_U8n1} are the extrapolations in DMET which scale as $1/\sqrt{N_c}$.  Due to the restricted small system sizes in DMET and large $U$, the resulting uncertainty is dominated by  the extrapolation.  The value, also shown in the main panel, is in good agreement with DMRG and FN, and only slightly outside error bars of the AFQMC result.

In Fig.~\ref{fig:extrap_DCA_DIAG} we show data for two DiagMC methods at $U/t=4$, $T/t=0.25$ and $n=0.8$, along with DCA data.  DiagMC  is done directly in the thermodynamic limit, and the results  become successively more precise as more and more expansion orders are added to the series. The results from the two diagrammatic series we show agree within error bars, with the $G^2\Gamma$ series converging more smoothly than $[G^{(0)}]^2\Gamma^{(0)}$. The convergence with expansion order in the regimes we present is very rapid, so that the value at order $\alpha=6$ or $7$ can be taken as representative for the infinite order series, with error bars estimated by statistics and by comparison to the results at the second largest order; in other words, extrapolation to $\alpha\rightarrow \infty$ is not needed. 

DCA for the 2D Hubbard model approaches the thermodynamic limit $\sim 1/N_c$. However, in the parameter regime considered here the many-body physics is converged with respect to $N_c$ and deviations from the thermodynamic limit are dominated by single-particle shell effects. In other regimes, especially at larger $U$, extrapolation in $1/N_c$ is required, see for example Refs.~\onlinecite{fuchs:2011,gull:2011:smu,leblanc:2013}.

The key result of this section is that in many cases extrapolation to the infinite system size limit is needed to obtain accurate results, with the value obtained by extrapolation significantly different from the value obtained by the largest size studied. For this reason we will typically display below both the extrapolated thermodynamic limit results and the finite size results that produced the extrapolation. 

\subsection{Sources of Uncertainty}
For clarity we repeat the main sources of uncertainties and the meaning of the error bars shown in the graphs for each technique; further details can be found in the sections on each method. 

$\bullet$ AFQMC: at $n=1$ error bars include all sources of uncertainty; stochastic errors and extrapolation to TL. For $n\neq1$, uncertainty from the constrained path approximation is not estimated by the error bar. 

$\bullet$ FN: error bars account for stochastic Monte Carlo errors and for extrapolation to the TL.  Uncertainties due to the fixed node approximation are not included in the error bar.

$\bullet$ MRPHF: results are not extrapolated to the TL and, on each finite system, an estimate of the uncertainty due to truncation in the number of Slater determinants is not included.

$\bullet$ UCCSD: error bars do not include an estimate of uncertainty for truncation of excitation order to doubles.

$\bullet$ DMRG: error bars include all sources of uncertainty; the extrapolation in the number of basis states and extrapolation to TL.

$\bullet$ DMET: error bars include all sources of uncertainty; uncertainty due to extrapolation in of the number of basis states of the impurity solver and extrapolation to TL as well as estimates of DMET self consistency convergence.

$\bullet$ DCA: error bars include all sources of uncertainty;  stochastic Monte Carlo uncertainties and an additional estimate of uncertainty due to extrapolation to the TL.

$\bullet$ DF: values are presented without error bars, the effect of neglecting non-ladder and higher order diagrams is not quantified.

$\bullet$ DiagMC: error bars include all sources of uncertainty; the stochastic Monte Carlo uncertainty at each expansion order and estimate of uncertainty in convergence of expansion order.

\section{Results at Intermediate to Strong Interaction Strength}\label{results_intermediate_strong}\label{sec:strong}

We begin our discussion of results with an analysis of an intermediate-to-strongly coupled case, namely $U/t=8$.  Throughout all figures we use  common legends, distinguishing techniques by symbol and color. We present both results for the thermodynamic limit and the finite system size data from which the thermodynamic limit results were obtained.  This information is useful in assessing both the importance of the extrapolation and other aspects of the performance of the method.  We use open symbols to denote values in the thermodynamic limit and filled symbols for finite size values from which the extrapolations are obtained.

\subsection{Half-Filled, particle-hole symmetric case ($U/t=8$, $n=1$, $t^\prime/t=0$) }\label{hfph_sec}

\begin{figure}[tbh]
\includegraphics[width=\linewidth]{{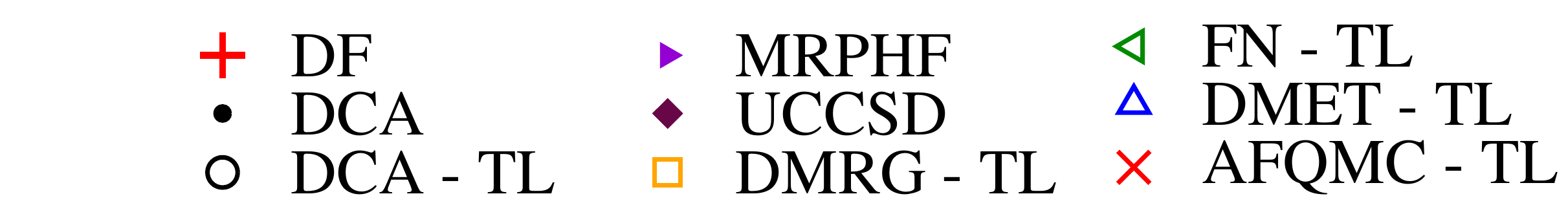}}
\includegraphics[width=\columnwidth]{{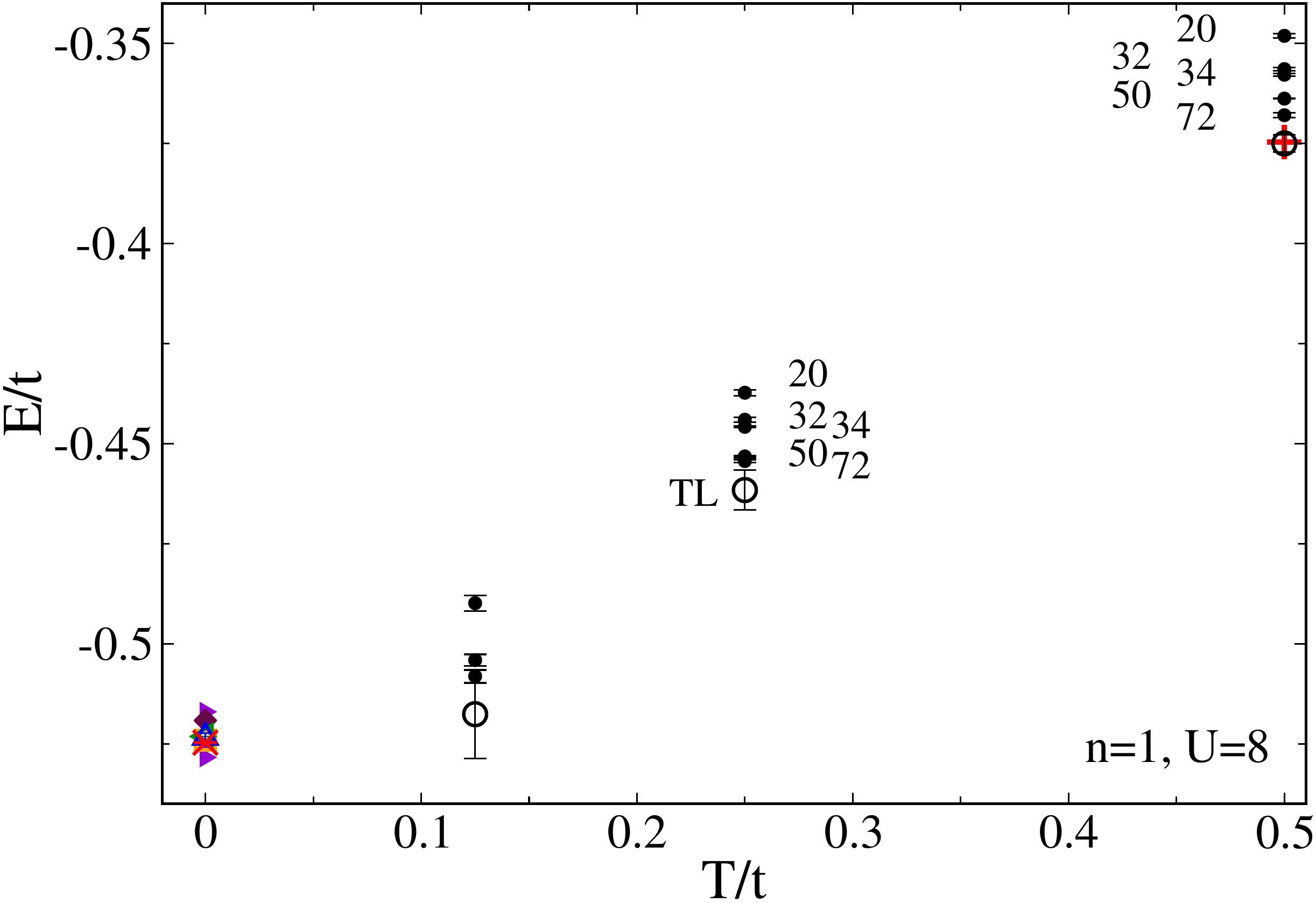}}
\caption{\label{fig:U8_n1_E_finiteT} (Color online)  Temperature dependence of the energy for $n=1$ for $U/t=8$ obtained by DCA (black circles) and DF (red cross) and compared to zero temperature results compiled from various techniques.  Solid symbols represent finite systems, open symbols represent extrapolations to the thermodynamic limit (TL).  }
\end{figure}

\begin{figure}[tbh]
\includegraphics[width=\linewidth]{{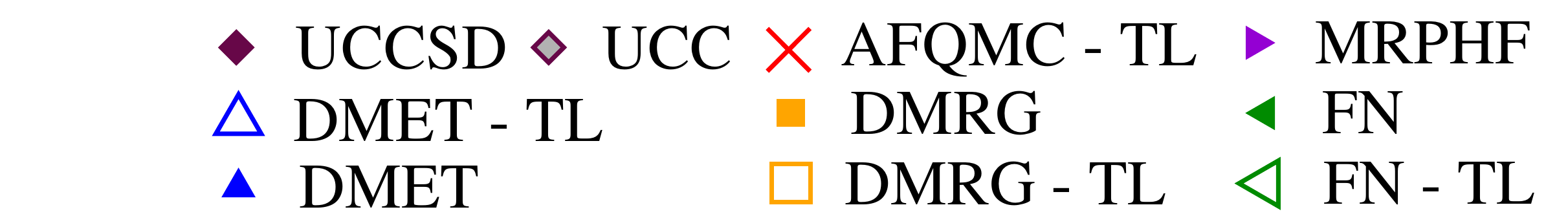}}
\includegraphics[width=\linewidth]{{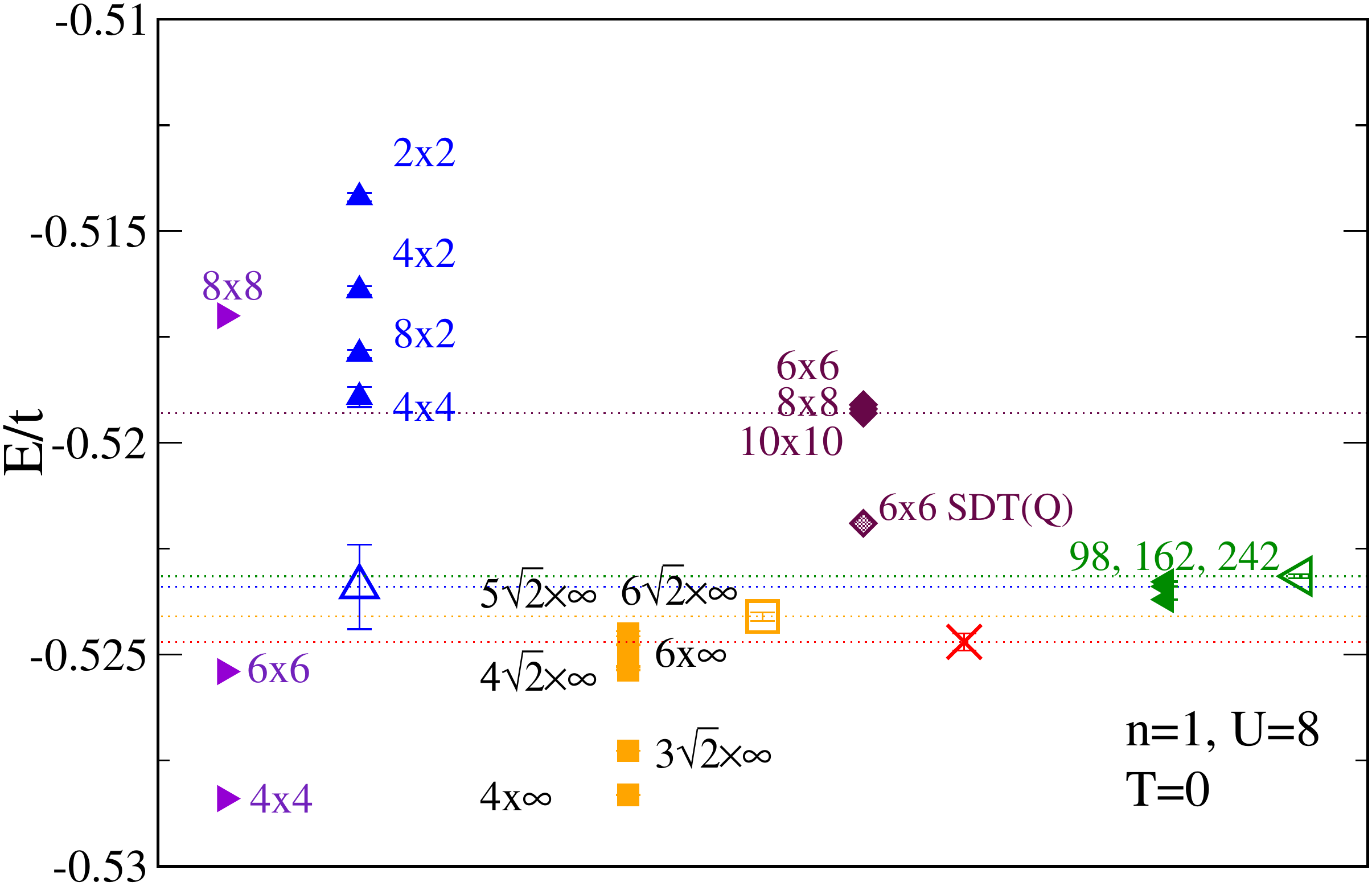}}
\caption{\label{fig:U8_n1_E_zeroT} (Color online) Thermodynamic limit (TL) ground state energy for $n=1$ for $U/t=8$ as obtained by various algorithms (open symbols). Also shown are the finite size systems (filled symbols with adjacent labels) from which the TL ground state energy was obtained. Data from AFQMC (red crosses), DMET (blue triangles), UCCSD (maroon diamonds), MRPHF (purple triangles), DMRG (orange squares), and FN (green triangles). Horizontal thin dotted lines show the best estimates for the ground state energy.}
\end{figure}

We begin our discussion with an analysis of the energy per site for a widely accessible parameter choice, half filling, showing in Fig.~\ref{fig:U8_n1_E_finiteT} the temperature dependence of the energy and in Fig.~\ref{fig:U8_n1_E_zeroT}  an expanded view of the $T=0$ energy. 

First we discuss the DCA results in Fig.\ref{fig:U8_n1_E_finiteT}; in this particle-hole symmetric parameter regime the impurity solvers have no sign problem and cubic complexity, meaning that reliable results can be obtained on relatively large clusters and that the Monte Carlo errors (here are on the order of $10^{-3}$) can be systematically reduced with additional computation. Thermodynamic limit data are obtained from the $1/N_c$ extrapolation.   Computational scaling towards low temperatures results in an increase of uncertainty for fixed computational time, and this is reflected in the uncertainty in the extrapolated values.  At $T/t=0.5$ our results agree within error bars with  high temperature series and lattice Monte Carlo data (see Ref. \onlinecite{leblanc:2013}).

The results of a DF calculation are shown at $T/t=0.5$ (lower $T$ data are not available). The DF technique neglects vertex functions of higher order than two-particle vertices.  Further, at the two-particle level we  sum only a ladder series in the spin and charge channels.    Despite these approximations, we see that the DF technique provides an energy which falls on top of that of the extrapolated thermodynamic limit DCA result.

Results from a variety of algorithms are available at zero temperature.  Fig.~\ref{fig:U8_n1_E_zeroT} presents an expanded view of the $T=0$ results, with the energy on the vertical axis and data for each method offset in the $x$-axis. Note that in some cases the thermodynamic limit results are further offset for clarity. 

We start our discussion of zero-$T$ results with a Monte Carlo technique, AFQMC, which is extrapolated to the thermodynamic limit.  In this case, finite size results are averaged over twisted-boundary conditions, which allows a smooth and rapid convergence to the thermodynamic limit. These results, obtained at half filling from Monte Carlo, are unbiased and therefore expected to be exact within a quoted uncertainty of  $\pm 0.0002t$.

DMRG results on cylinders of infinite length but finite width of $3$, $4$, $5$, and $6$ for $45$ degree rotated systems and width of $4$ and $6$ for non-rotated systems are shown. All the finite size data are after phase averaging, showing only very weak finite-size effects, so that an extrapolation to the thermodynamic limit  is feasible.
In this case, the estimation of uncertainty (as discussed in Sec.~\ref{sec:DMRG}) contains the uncertainty of each extrapolation and the
difference between the two orientations (rotated and non-rotated), both of which are on the order of $10^{-4}t$. The resulting energy is close to, but slightly outside of, the AFQMC results.
This extrapolation issue was discussed in more detail in Sec.~\ref{sec:extrapol}.

For DMET  we show results  obtained for finite clusters of size  $2\times 2$, $4\times2$, $8\times2$, and $4\times4$. The thermodynamic limit is obtained by extrapolating the $2\times 2$, $4\times 2$, and $4\times 4$ clusters in $1/\sqrt{N_c}$. Errors from the
solution of the finite impurity are  on the order of $10^{-4}t$. 
DMET cluster size convergence is slower at large $U$, thus $U/t=8$ corresponds
to the largest half filling DMET error bar discussed here.
The total thermodynamic limit uncertainty is estimated to be $0.001t$, and comes entirely 
from the thermodynamic limit extrapolation. The lower end of the DMET error bar lies at the average of 
the DMRG and AFQMC estimates.


For the FN technique, a diffusion Monte Carlo calculation based on the nodal structure of a trial wave function obtained with variational Monte Carlo, we show finite size results for a sequence of 45-degree rotated
    clusters with size 98, 162, and 242, which have the property of being
    closed shells at $U=0$. The results show only weak size dependence, so that the thermodynamic limit value shown is  close to the finite size results. However, the results are systematically above the values obtained by AFQMC and DMRG, while they are consistent with DMET.  This is a consequence of the fixed node approximation, which in this particular case resulted in a fixed node error of about $0.0015t$.

The results of UCCSD are shown for systems of size $6\times6$, $8\times8$, and $10\times10$, and exhibit weak finite size effects at this $U$ value. We see that the result is accurate to roughly the 1\% level. The deviation is caused by correlations that are not captured by singles and 
doubles. Higher order excitations (triples, quadruples, etc) will eventually recover the remaining energy.
To support this claim we show a single case in Fig.~\ref{fig:U8_n1_E_zeroT} labeled as UCCSDT(Q) which includes all triples and a subset of quadruples.  In this higher order approximation the deviation from other techniques is reduced by a factor of 2.  The higher order corrections are more important for these coupled cluster calculations than extrapolations in cluster geometry size, however, the improvement with increased excitation order converges slowly.  Also shown are approximate results including quadruples from small system sizes ($4\times 4$) which we label UCCSDTQ*.  While not exact, this approximation scheme produces a results which deviates from AFQMC by only $0.15\%$.

MRPHF calculations have been performed for several finite systems ($4\times 4$, $6\times 6$, and $8\times 8$) . As summarized in the methods description, reaching a constant level of accuracy would require a successively larger MR expansion. Results for larger systems are therefore solved less precisely; in particular the energy of the $8 \times 8$ lattice in Fig.~\ref{fig:U8_n1_E_zeroT} is too high. 
More sophisticated implementations and additional optimizations may make it possible to reach the accuracy needed to perform extrapolations to the thermodynamic limit.

\begin{figure}[tbh]
\includegraphics[width=\linewidth]{{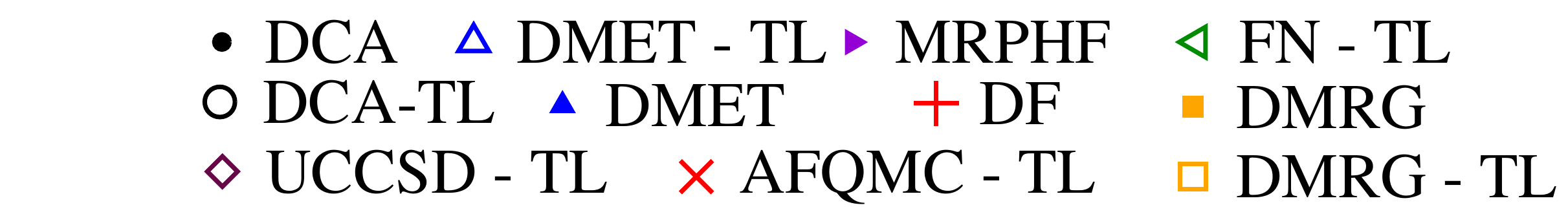}}
\includegraphics[width=\linewidth]{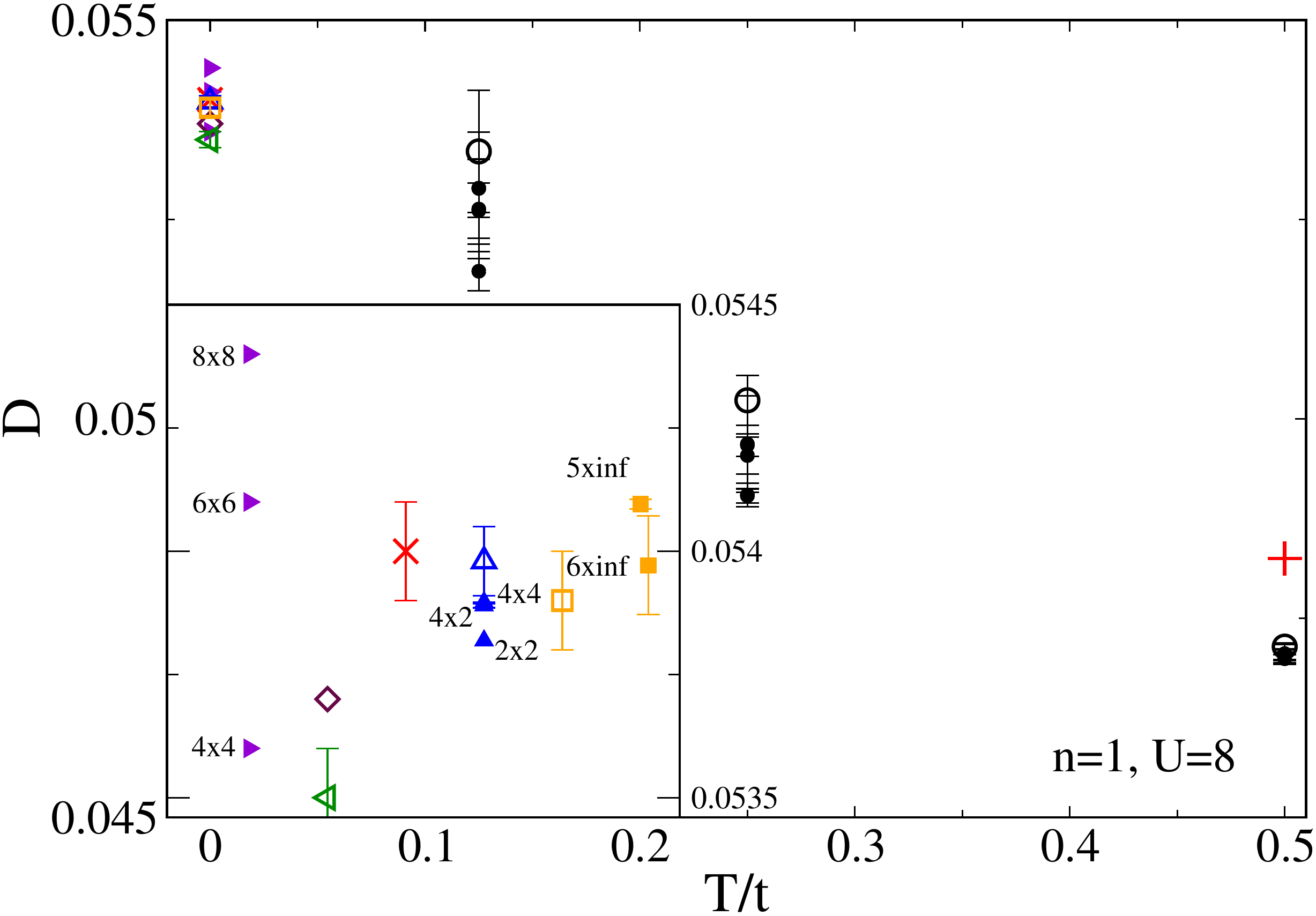}
\caption{\label{fig:U8_n1_D} (Color online)  Double occupancy data for $U/t=8$ and $n=1$. Main panel: temperature dependence of double occupancy, obtained from DCA (finite $T$, black circles) and DF (finite $T$, red plus sign), and the $T=0$ techniques AFQMC (red crosses), DMET (blue triangles), UCCSD (maroon diamonds), MRPHF (purple triangles), DMRG (orange squares), and FN (green triangles).  Solid symbols represent finite systems, open symbols represent extrapolations to the thermodynamic limit (TL).
Inset: data at $T=0$ reproduced with an arbitrary x-axis offset, from MRPHF, UCCSD, FN, DMET, DMRG, and AFQMC.}
\end{figure}

We now  discuss the results for the double occupancy in Fig.~\ref{fig:U8_n1_D} at $U/t=8$ and $n=1$.
Open symbols denote results in the thermodynamic limit, filled symbols results on finite systems. The finite-$T$ DCA results show that the double occupancy contribution rises as the temperature is lowered. The finite-$T$ results are consistent with the $T=0$ values obtained by AFQMC, DMRG, DMET, FN, and MRPHF.  At $T/t=0.5$, the double occupancy obtained from the DF technique is also shown.  Unlike the total energy, the DF double-occupancy shows deviations from the DCA result,  suggesting a cancellation of errors in the kinetic and potential energy terms. As for all other points we have examined, the DF method produces results which lie between single-site DMFT values (not shown) and the extrapolated DCA results.

The inset shows the various $T=0$ values. Within error bars, there is agreement between AFQMC, DMET and DMRG results for the double occupancy.
DMET obtains a value (after thermodynamic limit extrapolation) comparable to AFQMC and, overall, shows a weaker system size dependence than for the energy.

UCCSD and FN produce a double occupancy which is underestimated as compared to AFQMC and DMET. Finite size effects of FN are on the order of $0.0001$.
Finally, for MRPHF we quote two values for $4\times 4$ and $6\times6$ systems which show a system size dependence on the order of $0.001$.  This makes a thermodynamic limit extrapolation impractical. 

\subsection{Doped strongly correlated regime ($U/t=8$, $n=0.875$, $t'/t=0$)}
\begin{figure}[tbh]
\includegraphics[width=\linewidth]{{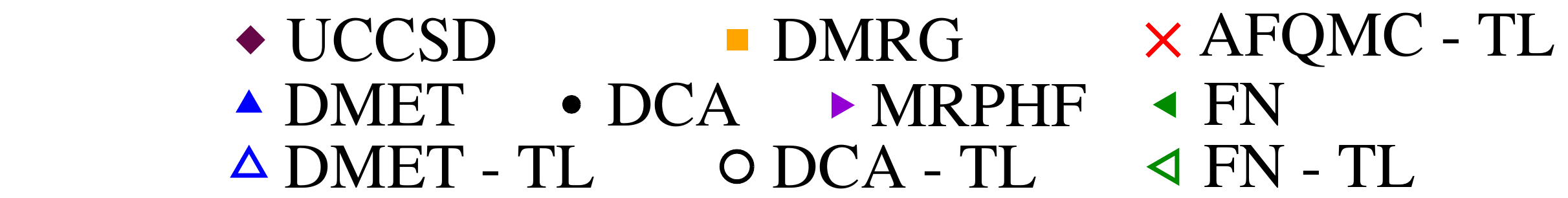}}
\includegraphics[width=\linewidth]{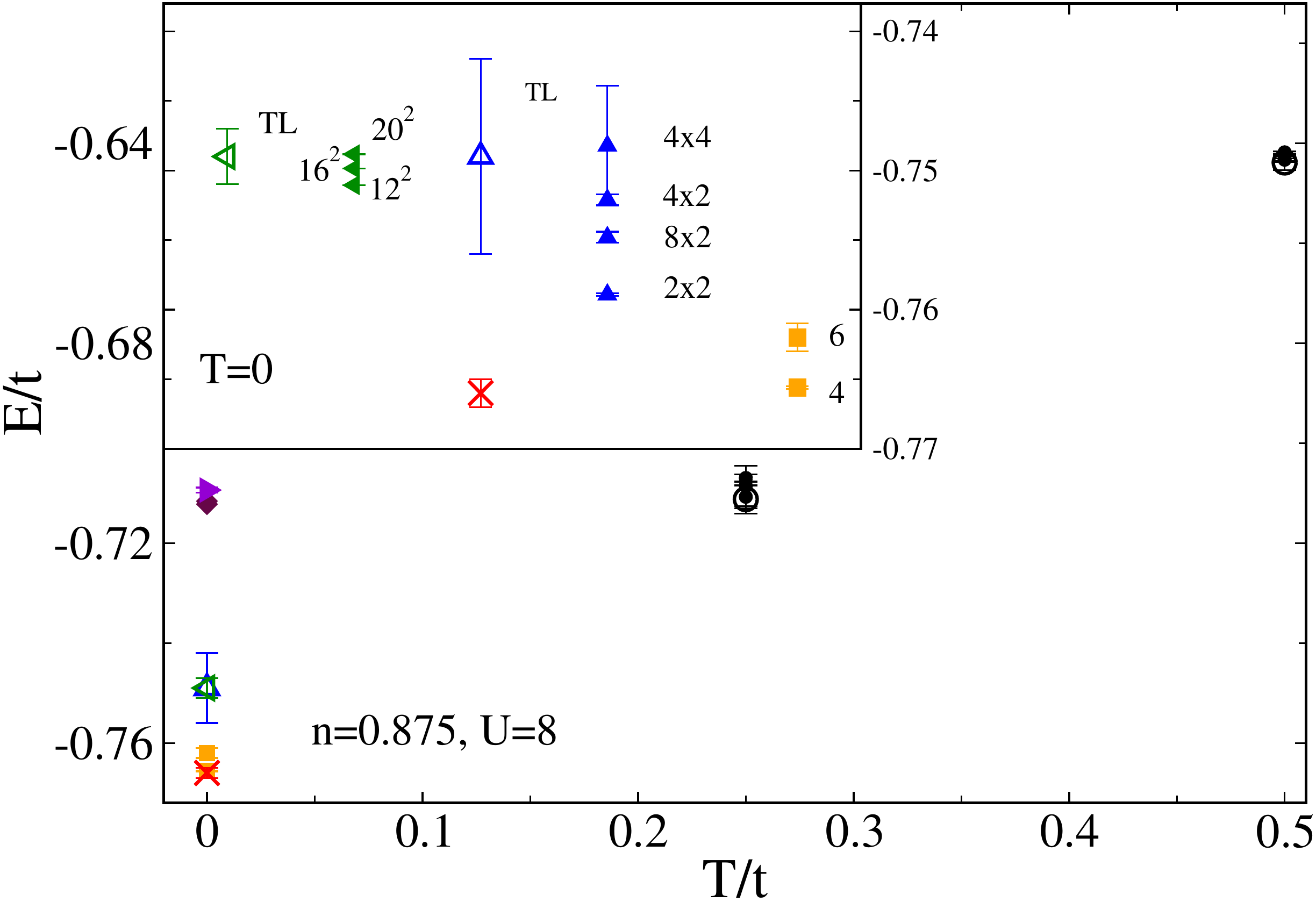}
\caption{\label{fig:U8_np875_E}  (Color online)  Data for $n=0.875$ for $U/t=8$. Main panel: temperature dependence of $E/t$ compiled from various techniques.  Solid symbols represent finite systems, open symbols represent extrapolations to the thermodynamic limit (TL).  Finite $T$ results are shown for DCA (black circles), and  zero-$T$ data from AFQMC (red crosses), DMET (blue triangles), UCCSD (maroon diamonds),  DMRG (orange squares), and FN (green triangles).
Top left inset: zoom in to the zero-$T$  data from DMET, AFQMC, FN, and DMRG including finite system size data (as labeled). }
\end{figure}

\begin{figure}[tbh]
\includegraphics[width=\linewidth]{{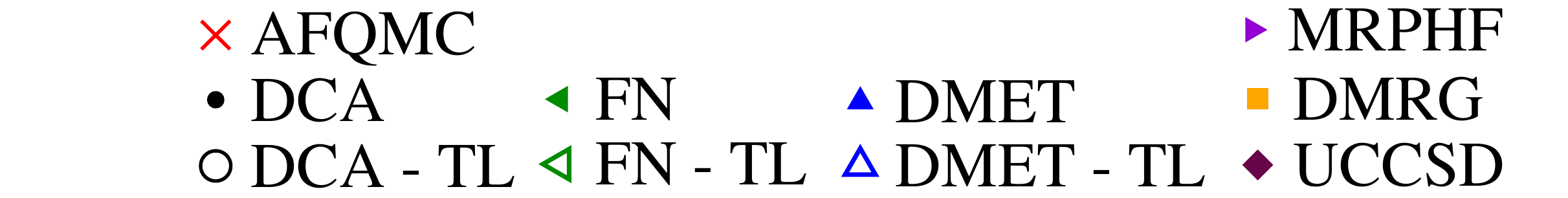}}
\includegraphics[width=\linewidth]{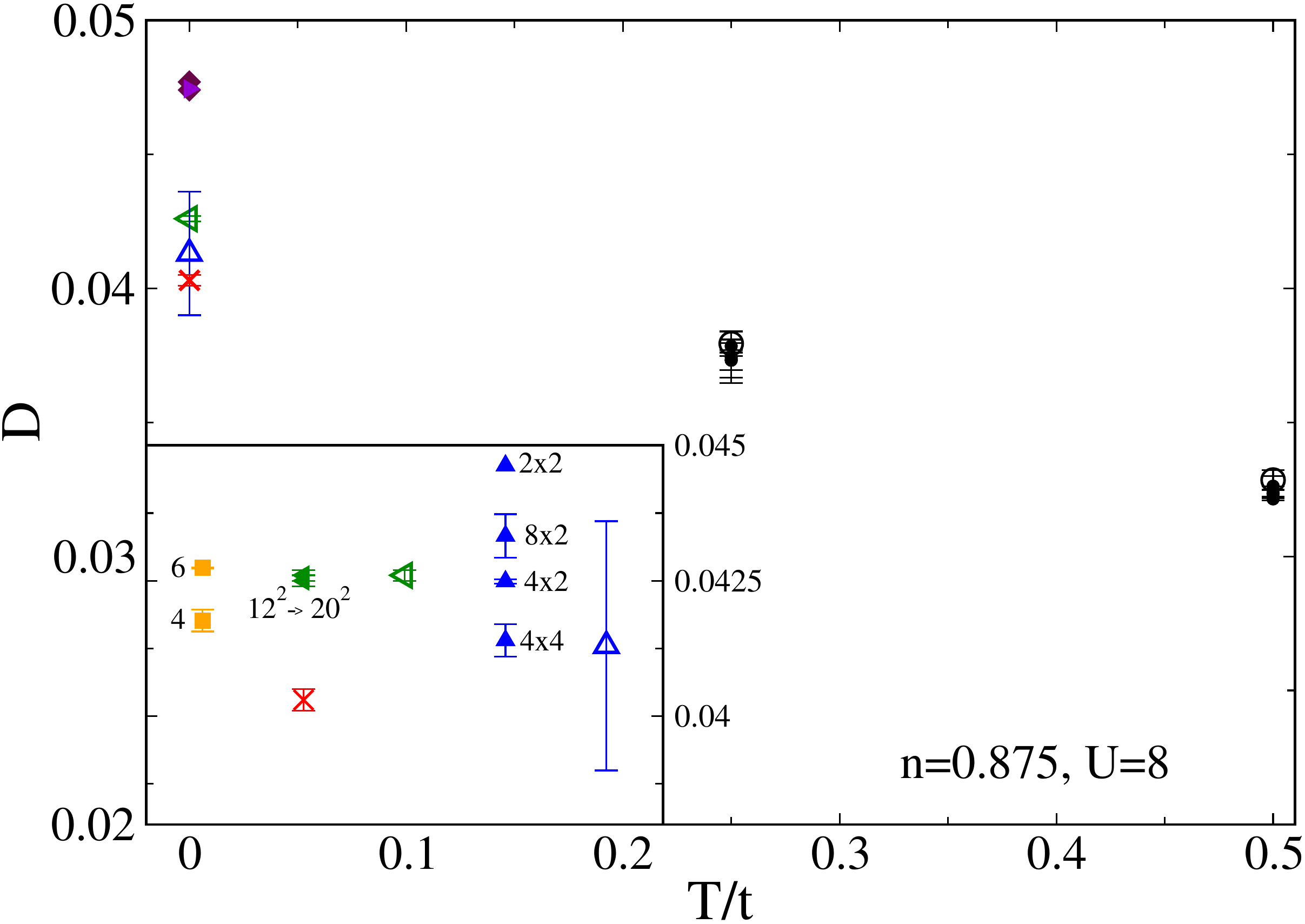}
\caption{\label{fig:U8_np875_D} (Color online) Data for $n=0.875$ for $U/t=8$. Main panel: temperature dependence of double occupancy, $D$, compiled from various techniques.  Solid symbols represent finite systems, open symbols represent extrapolations to the thermodynamic limit (TL).  Finite $T$ results are shown for DCA (black circles), and  zero-$T$ data from DMET (blue triangles), MRPHF (purple triangles), UCCSD (maroon diamonds), DMRG (orange squares), and FN (green triangles).
Inset: zoom in to the zero-$T$  data from  FN, AFQMC, DMET, and DMRG.}
\end{figure}

The half-filled particle-hole symmetric case of Sec.~\ref{hfph_sec} is in many ways ideally suited for numerical algorithms: a large charge gap allows methods like the DMRG to quickly converge, and particle-hole symmetry makes Monte Carlo simulations without a sign problem possible. We now turn to a case which is particularly difficult to simulate, where we expect results to be substantially less accurate than for the half filled case. This parameter regime shows metallic behavior, strong particle-hole asymmetry, and interesting inhomogeneous phases in the ground state. In Fig.~\ref{fig:U8_np875_E} we plot the total energy per site and in Fig.~\ref{fig:U8_np875_D} the double occupancy per site at  $U/t=8$ and $ n=0.875$.  

The main panel presents data as a function of temperature. The DCA results remain consistent with $T=0$, but  results are not available at the lowest benchmark temperature, $T/t=0.125$,  due to a large sign problem in the Monte Carlo impurity solver.  The finite size effects, at the system sizes accessible in DCA, are smaller than in the $n=1$ case.

The inset to Fig.~\ref{fig:U8_np875_E} presents data at $T=0$ with an arbitrary $x$-axis offset added for clarity. 
The AFQMC simulation, using a (non-variational) constrained path approximation in the absence of particle-hole symmetry, yields a result for the total energy that is lower than the one obtained from DMET, FN, and DMRG.
The total energy difference is $\sim$1\% when compared to finite sized DMRG, and $\sim 1.4\% (2.1\%)$ when compared to the DMET $8\times 2$ (thermodynamic limit) cluster respectively, and $\sim2.1\%$ in comparison to FN.

DMRG shows the results for cylinders of infinite length and finite widths of
$4$ and $6$ lattice sites after using phase averaging.
The energy is higher for the wider cylinder, and for the width $6$ cylinder the
energy is above the energy from AFQMC.  
Given that the extrapolation is performed with only two widths, we consider the extrapolated DMRG value to be not reliable in this case and omit it entirely.

DMET shows a large system size dependence and a dependence of the thermodynamic limit value on the cluster sequence chosen for the extrapolation. We show an extrapolation based on $2\times2$, $4\times 2$, and $8\times 2$ clusters. The use of the $8\times 2$ cluster
allows inhomogeneous order to develop, giving an extrapolated 
value of $E/t=-0.749(7)$. The extrapolation using the $4\times 4$ rather than $8\times 2$ 
cluster, which does not allow for inhomogeneous order, yields a value of $E/t=-0.737(5)$. 
Since the energy changes non-monotonically: the $8 \times 2$ energy lies above the $2 \times 2$ energy, but below the $4\times2$ energy, the uncertainty in the thermodynamic
limit extrapolation is very large, and does not provide any more information
than the results obtained from the largest clusters. 

The FN method shows a clear finite system size dependence. The infinite system value is estimated from the $16\times 16$ and $20 \times 20$ 
values, and finite size errors are on the order of $0.001t$, much larger than the stochastic errors of $0.00001t$.  Here, the 
FN results are consistent with DMET extrapolation which omits the $4 \times 4$ cluster and these are considerably higher 
than AFQMC.  This is suggestive of a fixed node error of $\approx 0.015t$, indicating that a uniform variational wave 
function may not be enough to fully account for the nature of the ground state.
Indeed, the VMC error is of the order of $0.022t$, much larger than the one obtained at half filling which was $\approx 0.004t$. In both cases,
the FN projection improves the VMC results by the same order of magnitude.

UCCSD and MRPHF results are much higher in energy ($E/t=-0.7094(5)$ for MRPHF ($16 \times 4$ system) and $E/t = -0.7122$ for UCCSD, barely visible on the main panel), an indication that correlated metallic states are difficult to capture with these methods.  Although not shown, data is available for UCCSD(T) (perturbative inclusion of triples) in the supplemental material,\cite{suppl} which improves upon the value from UCCSD, and gives $E/t = -0.7272$ ($-0.7281$) for a $16\times 4$ ($16\times 8$)
cluster. Full inclusion of triples (UCCSDT) lowers the
$16\times 4$ estimate to $E/t = -0.7427$.  The MRPHF results indicate the need for a much larger MR expansion than that afforded in this work.

In this parameter regime, ordered `stripe' phases might exist. 
However, the precise form of these stripes is strongly influenced by choice of finite size systems (e.g. width and orientation of the cylinder in DMRG and shape of the cluster in DMET) that are used for the 
thermodynamic extrapolation and the approximations used to solve that finite system. The finite temperature algorithms have not 
reached the onset of inhomogeneous states at the lowest temperature accessible. The precise nature of the inhomogeneities in the ground state in this parameter regime is still open.

We finally briefly mention the results for double
occupancy in Fig.~\ref{fig:U8_np875_D}  for $U/t=8$ and $n=0.875$. As was the case for the energies, the finite-$T$ results smoothly connect to the zero-$T$ values. MRPHF and UCCSD overestimate the double occupancy by close to $15\%$. The remaining ground state methods (DMRG, AFQMC, DMET, and FN) present consistent values in the range between $0.04$ and $0.043$. Both FN and AFQMC values contain additional (fixed node and constrained path) errors that are not estimated by the error bar.

\subsection{Half-filled, non particle-hole symmetric case $(U/t=8,n=1,t'/t=-0.2)$}
\begin{figure}[bth]
\includegraphics[width=\linewidth]{{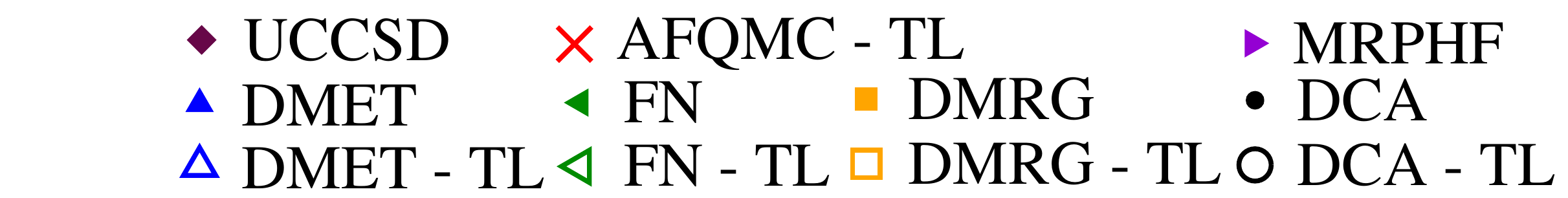}}
\includegraphics[width=\linewidth]{{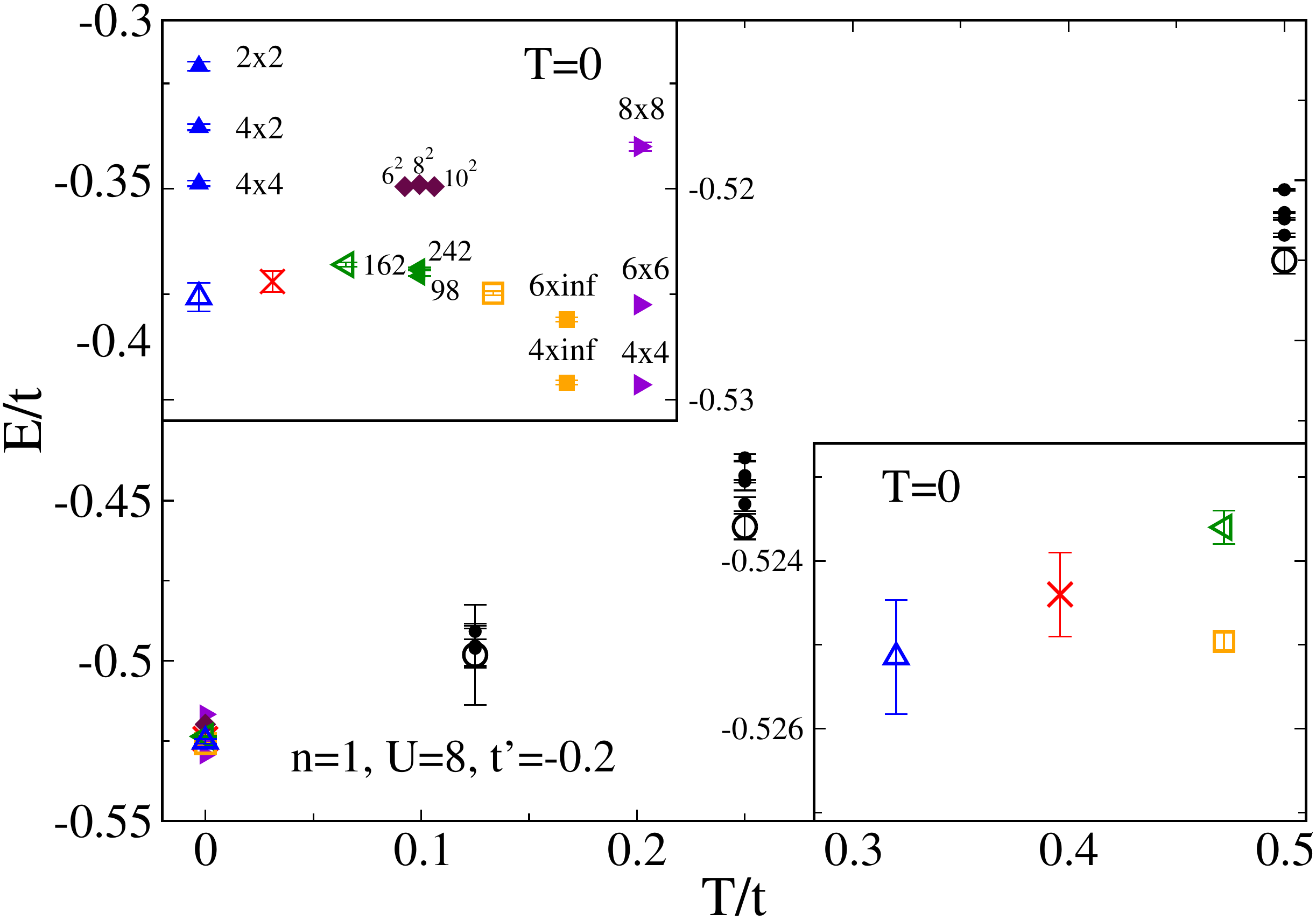}}
\caption{\label{fig:U8_n1_tp-0.2_E}(Color online)  Data for $n=1$ for $U/t=8$ with $t^\prime/t=-0.2$. Main panel: temperature dependence of $E/t$ compiled from various techniques.  Solid symbols represent finite systems, open symbols represent extrapolations to the thermodynamic limit (TL).  Finite $T$ results are shown for DCA (black circles), and  zero-$T$ data from AFQMC (red crosses), DMET (blue triangles), UCCSD (maroon diamonds), MRPHF (purple triangles), DMRG (orange squares), and FN (green triangles).
Top left inset: zoom in to the zero-$T$  data from MRPHF, UCCSD, DMET, FN, DMRG, and AFQMC, including finite system size data (as labeled) for MRPHF, FN,  DMET, and DMRG. Bottom right panel: enlarged region of the top left inset showing DMET, DMRG, FN, and AFQMC data at $T=0$, including error bars, extrapolated to the infinite system size.}
\end{figure}
We now turn our attention to a case of half filling without particle-hole symmetry, by adding a second nearest neighbor hopping $t^\prime$.  An overview of the energies from several algorithms for $U/t=8$, $n=1$ and $t^{\prime}=-0.2$ is shown in Fig.~\ref{fig:U8_n1_tp-0.2_E}.  

The main panel shows the temperature-dependence of the data. The DCA results available at finite $T$ show almost no sign problem for $T/t=0.5$ and $T/t=0.25$, but are  hampered by a severe sign problem at $T/t=0.125$.  The results are consistent within error bars with the zero-temperature results.

As at $U/t=8$, $n=0.875$, $t'=0$, the AFQMC is approximate because of a constrained path approximation due to the lack of particle-hole symmetry. Despite this, the results are in agreement with both the DMET and DMRG results.

The DMET results  are obtained on clusters of size $2\times 2$, $2 \times 4$, and $4\times 4$. Errors of the individual finite size systems are substantially smaller than the system size dependence. The DMET  thermodynamic limit is consistent with the thermodynamic estimates obtained from DMRG (from cylinders of width $4$ and $6$) and  from AFQMC. This is even more evident in the bottom right inset, which displays the thermodynamic limit estimates on a smaller scale.

FN results are higher in energy than AFQMC and
DMET (well within two joint standard deviations) and are
higher than DMRG by 0.0013t. As seen in
previous plots, the finite system size dependence of the
fixed node results is small on this scale.


UCCSD results show only small finite size effects and an overall energy $\approx$1\% higher than other techniques.
The MRPHF results  obtained on finite systems show an energy that rises rapidly as the system size is increased. As in the case of $t'=0$, a systematic extrapolation to the thermodynamic limit is not possible, and we only present results on finite systems.

\begin{figure}
\includegraphics[width=\linewidth]{{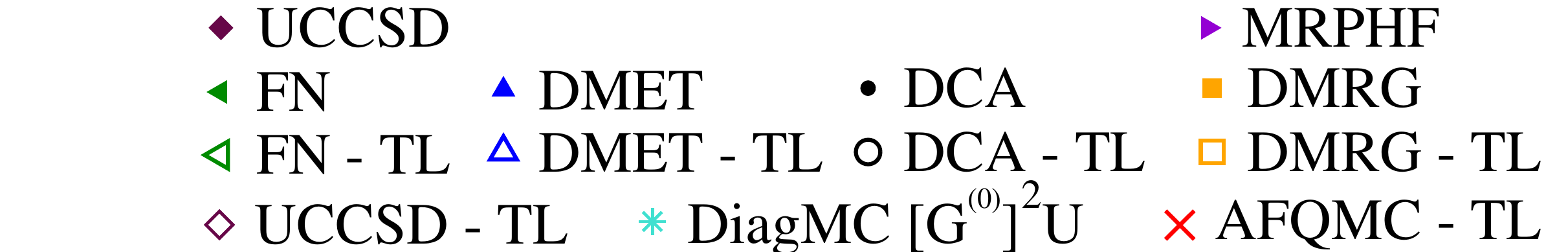}}
\includegraphics[width=\linewidth]{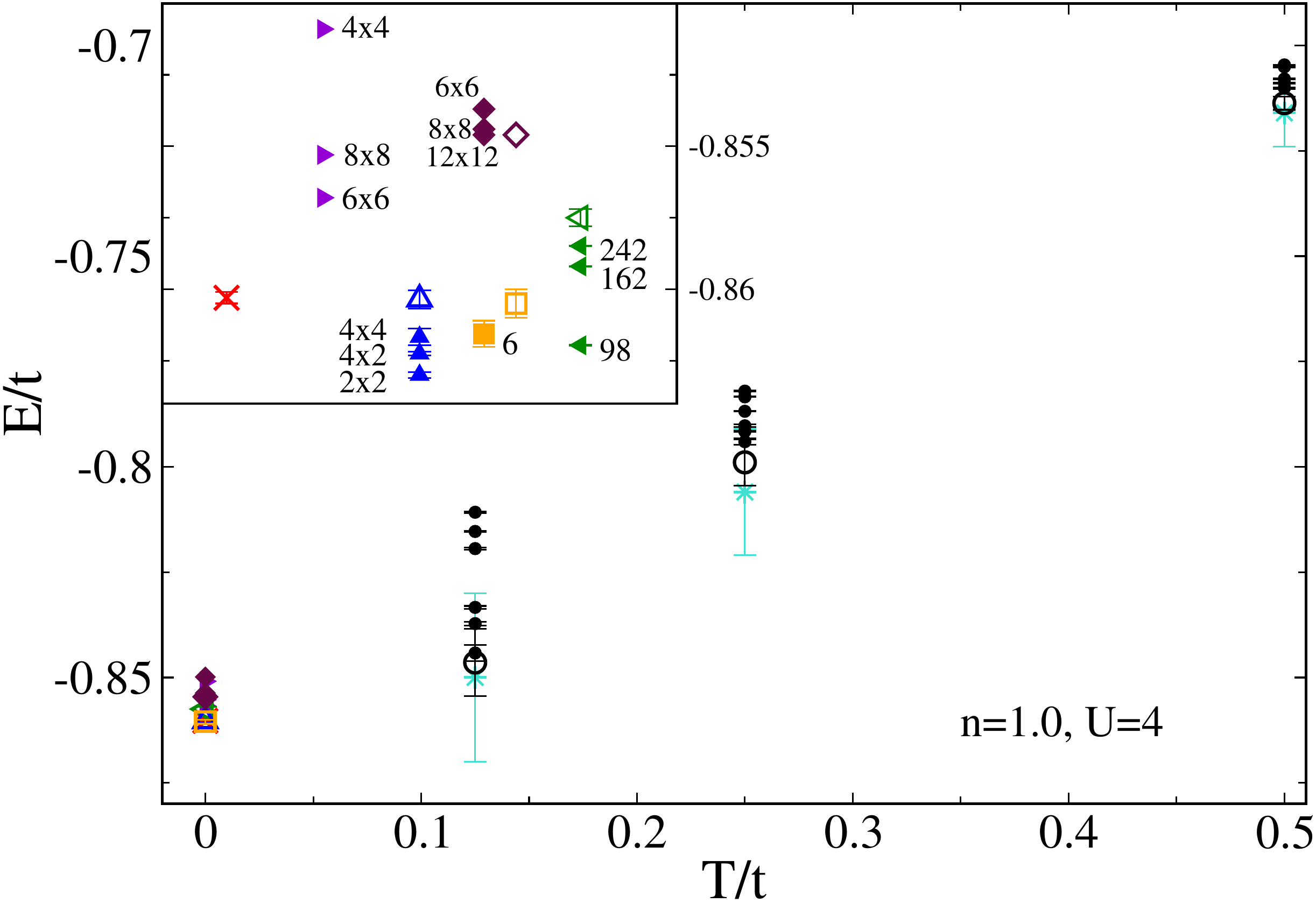}
\caption{\label{fig:U4_n1_E}(Color online)  Data for $n=1$ for $U/t=4$. Main panel: temperature dependence of $E/t$ compiled from various techniques.  Solid symbols represent finite systems, open symbols represent extrapolations to the thermodynamic limit (TL).  Finite $T$ results are shown for DCA (black circles) and DiagMC (turquoise stars), and  zero-$T$ data from AFQMC (red crosses), DMET (blue triangles), UCCSD (maroon diamonds), MRPHF (purple triangles), DMRG (orange squares), and FN (green triangles).
Top left inset: zoom in to the zero-$T$  data from MRPHF, DMET, FN, DMRG, and AFQMC, including finite system size data (as labeled) for MRPHF, FN, DMRG, and DMET.}
\end{figure}

\begin{figure}
\includegraphics[width=\linewidth]{{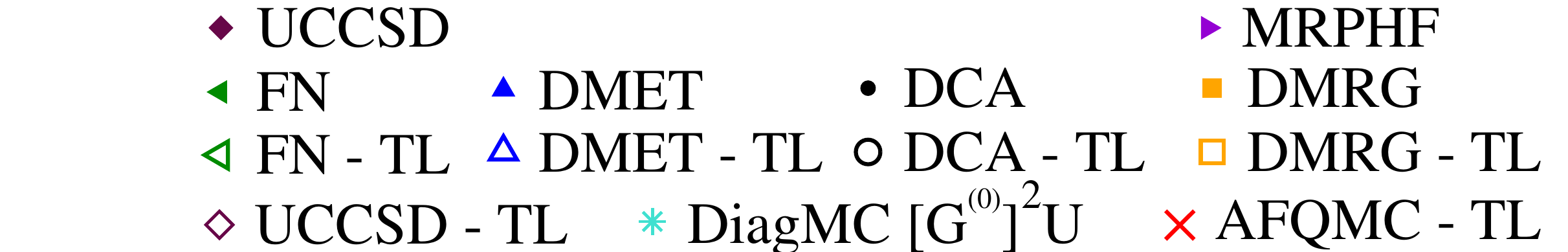}}
\includegraphics[width=\linewidth]{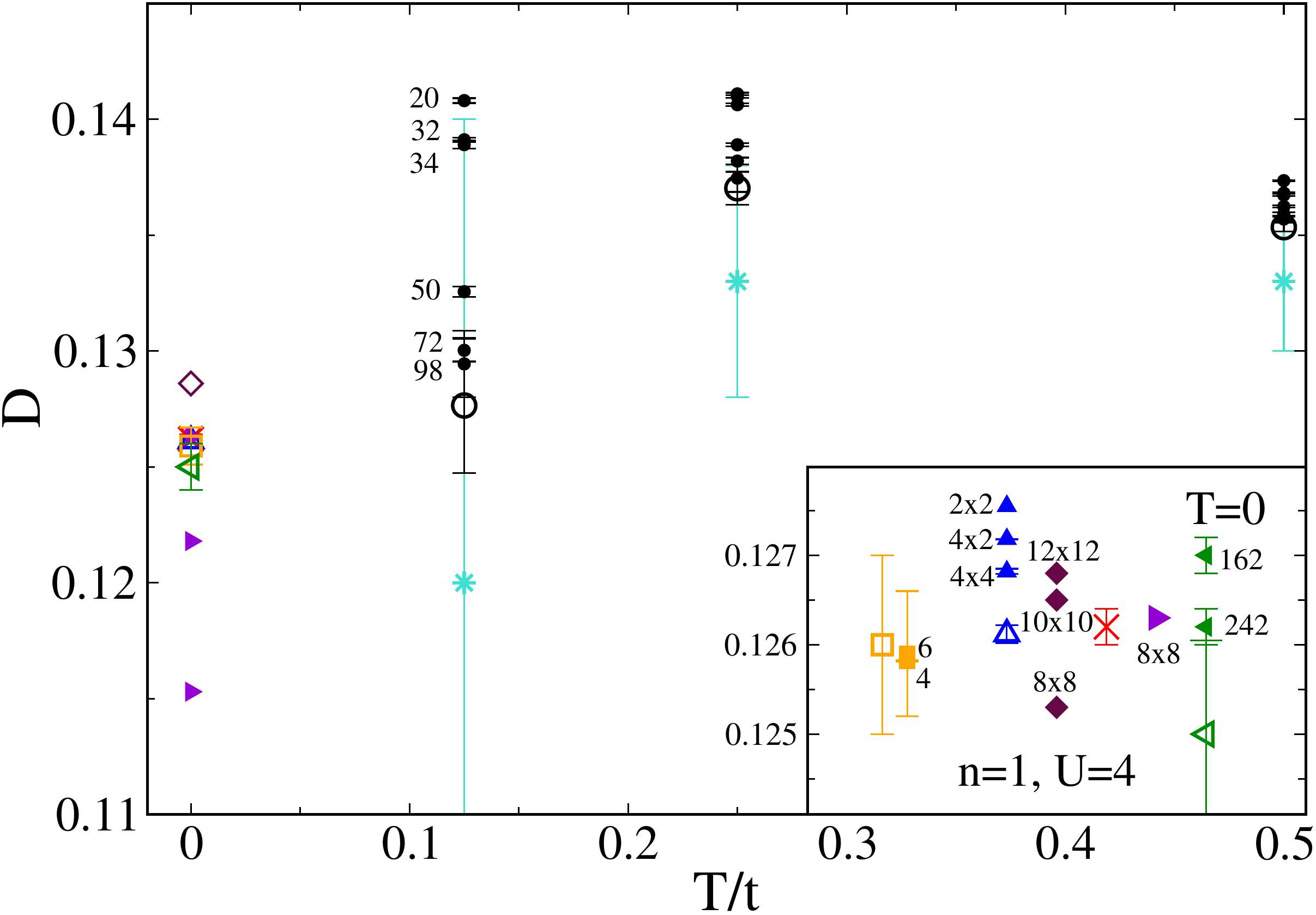}
\caption{\label{fig:U4_n1_D} Data for $n=1$ for $U/t=4$. Main panel: temperature dependence of double occupancy, $D$, compiled from various techniques.  Solid symbols represent finite systems, open symbols represent extrapolations to the thermodynamic limit (TL).  Finite $T$ results are shown for DCA  and DiagMC (turquoise stars), and  zero-$T$ data from DMET (blue triangles), UCCSD (maroon diamonds), MRPHF (purple triangles), DMRG (orange squares), and FN (green triangles).
Inset: zoom in to the zero-$T$  data from  DMRG, FN, UCCSD (only finite systems data), DMET, and AFQMC. }
\end{figure}
\section{Results in the Intermediate coupling regime} \label{results_intermediate}\label{sec:intermediate}
In this section we repeat the previous discussion for an interaction strength of half the size, $U/t=4$. As before we start our discussion at half filling. We then discuss a correlated metallic case with $20\%$ doping.

\subsection{Half-Filled, particle-hole symmetric case $(U/t=4,n=1,t'/t=0)$}\label{sec:U4n1}

In Fig.~\ref{fig:U4_n1_E} we report the energy as a function of temperature.
At finite $T$ and $U/t=4$, both DCA and the diagrammatic Monte Carlo method for the $[G^{(0)}]^2U$ series provide results in the thermodynamic limit. DCA results in the thermodynamic limit  are extrapolated from finite clusters, DiagMC results  are extrapolated in the expansion order. The results are consistent within the error bars of the respective methods.
The large error bars of the extrapolation in DiagMC-$[G^{(0)}]^2U$ mainly come from a conservative estimate of the diagram-order extrapolation error. DCA shows surprisingly large finite size effects which persist above $N_c=72$, unlike at $U/t=8$.  While each individual $N_c$ result has uncertainties in the energy on the order of $10^{-4}t$, the spread in values results in large uncertainty when extrapolated to the thermodynamic limit.

We now move to the zero-temperature methods,
which are shown in the inset of Fig.~\ref{fig:U4_n1_E}.
 AFQMC provides numerically exact ground state energies for this system. The value quoted is $E/t= -0.8603(2)$, which is in agreement with the DMET value of $E/t = -0.8604(3)$.
 and the DMRG value of $E/t = -0.8605(5)$. 
DMET values are obtained from an extrapolation of $2\times2$, $2\times 4$, and $4\times 4$ clusters.
DMRG values are obtained from an extrapolation of widths $3$, $4$ and $5$ for $45$-degree rotated cylinders and of width $4$ and $6$ for non-rotated cylinders.

The results obtained by AFQMC, DMRG, and DMET are in excellent agreement with 
recent calculations obtained from linearized auxiliary fields Monte Carlo (LAQMC) available in the literature\cite{Sorella11}, which gives $E/t= -0.85996(5)$.
FN results are higher in energy ($E/t = -0.8575(3)$), and unlike in previous cases for stronger interaction, a clear dependence on the finite system studied is visible. 
The FN projection technique leads to an energy gain of $\approx 0.002t$ with respect to the VMC result of $E/t=-0.8558(5)$. This number can be compared with a previous estimation of the thermodynamic limit in VMC obtained with a slightly less accurate variational state, see
Ref.~\onlinecite{tahara:2008}. 

The UCCSD thermodynamic limit overestimates the energy by $\sim 0.7\%$.
MRPHF values, shown as purple right triangles in the main panel and inset, show large finite size effects and are higher than the values obtained with other methods. We see that as system size is increased, the energy increases rapidly.

In Fig.~\ref{fig:U4_n1_D} we report the double occupancy vs $T$.  At finite $T$, the DCA results show a clear rise in $D$ as $T$ decreases from $0.5t \to 0.25t$.  However, this trend reverses as T decreases further.   A similar behavior is obtained also by using DiagMC,
demonstrating that this is a genuine effect present in the Hubbard model.

These trends are consistent with the $T=0$ data, which lie below all of the DCA data points at finite $T$. For clarity of presentation we again display this data in the inset and add an arbitrary x-axis offset. We see that finite size effects in DMET are very small, and that the extrapolation of DMET agrees perfectly with AFQMC.  Finite sized FN results produce values comparable to  to DMRG, DMET and AFQMC. However, extrapolation in FN results in an underestimate of the double occupancy, although within uncertainty.
In the DMRG simulations, phase averaging has greatly reduced finite size effects, 
and the DMRG error bars are determined by the truncation errors.
 Within those error bars, DMRG results are consistent with AFQMC and DMET.


\subsection{Doped case $(U/t=4,n=0.8,t'/t=0)$}
\begin{figure}
\includegraphics[width=\linewidth]{{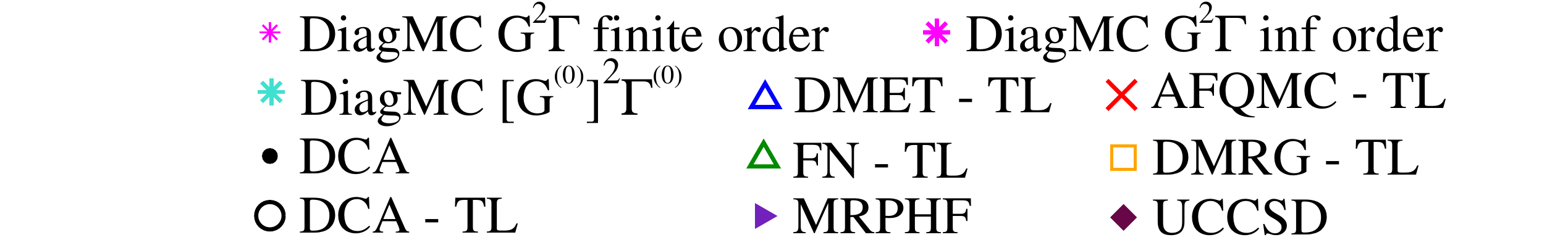}}
\includegraphics[width=\linewidth]{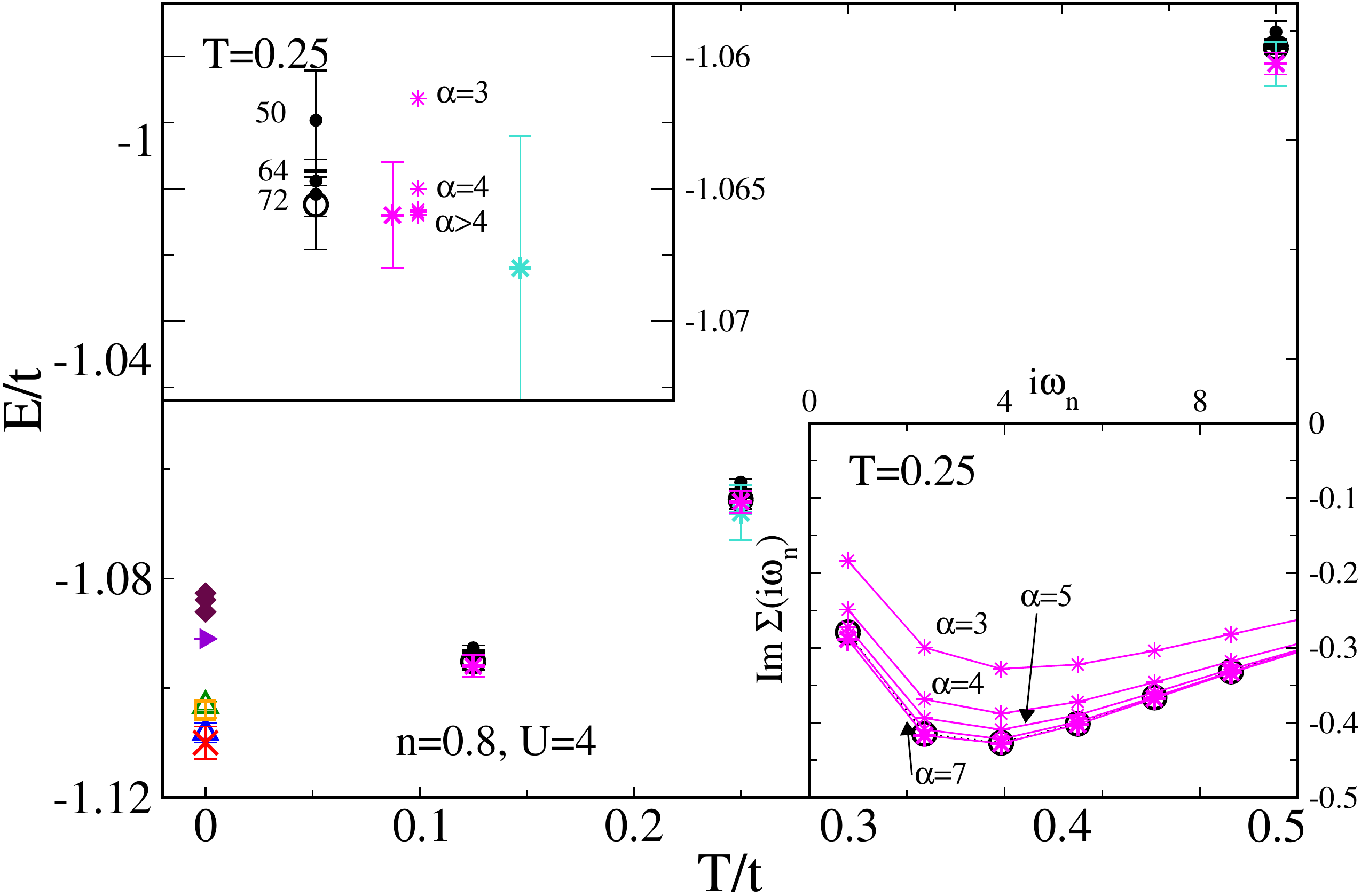}
\caption{\label{fig:U4_np8_E} (Color online)  Data for $n=0.8$ for $U/t=4$. Main panel: temperature dependence of $E/t$ compiled from various techniques.  Solid symbols represent finite systems, open symbols represent extrapolations to the thermodynamic limit (TL).  Finite $T$ results are shown for DCA (black circles) and DiagMC (pink and turquoise asterisks), and  zero-$T$ data from AFQMC (red crosses), DMRG (orange squares), FN (green triangles), and DMET (blue triangles).
Top left inset: zoom in to the $T/t=0.25$  data from DCA and two types of DiagMC for different orders $\alpha=3, 4, \dots$. Bottom right inset:  plot of the imaginary part of the local self energy ${\rm Im}\Sigma(i\omega_n)$ from DiagMC for different expansion orders $\alpha$ and from DCA (black circles covered by magenta stars). }
\end{figure}

Away from half filling (with $t'=0$), we can perform further comparisons at finite $T$ between DCA and DiagMC at $U/t=4$.  We begin the discussion with the inset of Fig.~\ref{fig:U4_np8_E}, which shows the convergence of the imaginary part of the local Matsubara self-energy of the $G^2\Gamma$ and $G^2W$  DiagMC series as a function of evaluation order. The values are compared to DCA results. We see that the first six orders of the series are precise enough to get good agreement of the Matsubara self-energy in the thermodynamic limit, and convergence is rapid. While deviations are visible in the energy, these are attributed to differences of the chemical potential, {\it i.e.} the real part of the self-energy. 

The top left inset of Fig.~\ref{fig:U4_np8_E} shows the convergence of the energy in DCA and two different series of DiagMC, $[G^{(0)}]^2U$ and $G^2\Gamma$, for increasing order of the diagrammatic resummations, $\alpha$. The values obtained from the three techniques are within error bars.

\begin{figure}
\includegraphics[width=\linewidth]{{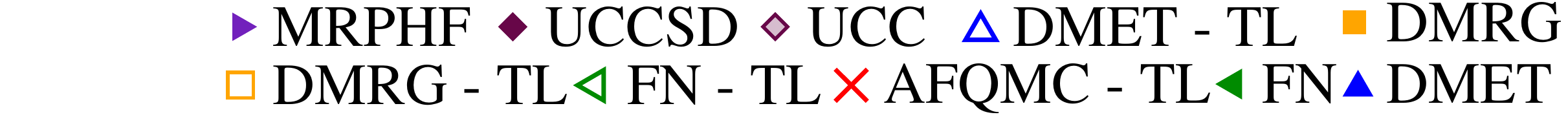}}
\includegraphics[width=\linewidth]{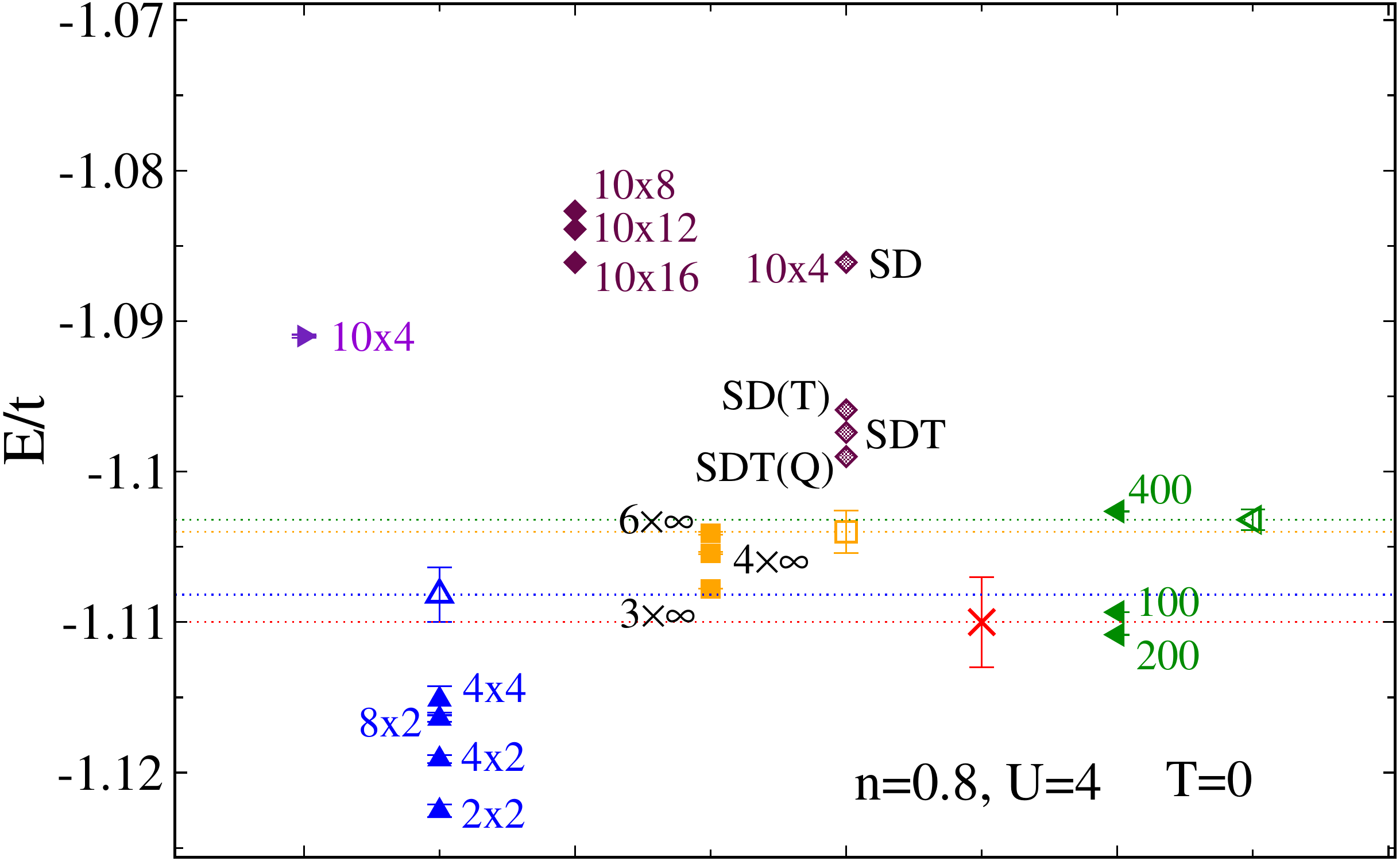}
\caption{\label{fig:U4_np8_E_Tzero} (Color online)  Thermodynamic limit (TL) ground state energy for $n=0.8$ for $U/t=4$  as obtained by various algorithms (open symbols). Also shown are the finite size systems (filled symbols with adjacent labels) from which the thermodynamic limit ground state energy was obtained. Data from AFQMC (red crosses), DMET (blue triangles), UCCSD (maroon diamonds), UCC on a $10\times 4$ lattice (shaded diamonds), MRPHF (purple triangles), DMRG (orange squares), and FN (green triangles). Horizontal thin dotted lines show the best estimates for the ground state energy.}
\end{figure}

The $T>0$ values smoothly connect to $T=0$ (although a precise comparison is not possible because we lack a quantitative functional form to extrapolate the $T>0$ values to $T=0$) which we display separately in Fig.~\ref{fig:U4_np8_E_Tzero}
 where data from DMET, AFQMC (constrained path), UCCSD, MRPHF, FN, and DMRG are shown.  In this case,  MRPHF and UCCSD are systematically higher in energy from the other techniques.  
In the case of UCCSD we see larger finite size effects than at $U/t=8$.  Inclusion of higher orders of excitation (perturbative triples (T), triples T and perturbative quadruples (Q)) suggests that the dominant error is associated with the truncation of the excitation order and not finite size effects. 
The FN result is in agreement with the value from DMRG.  At slightly lower energy AFQMC (constrained path) and DMET are in close agreement.  Overall, the spread in energies is similar to that shown away from half filling at $U/t=8$ (see Fig.~\ref{fig:U8_np875_E}) but smaller in magnitude, perhaps due to better convergence for weak coupling in some techniques.


\section{Results in the weak coupling regime}\label{results_weak}\label{sec:weak}
In the weak coupling limit we restrict the presentation of data to the half-filled case since the correlated metallic phase is not qualitatively distinct from $U/t=4$.  Data sets for doped, weakly correlated systems are available in the supplemental material.\cite{suppl}
\subsection{Half-Filled, particle-hole symmetric case ($U/t=2$, $n=1$, $t^\prime/t=0$) }
\begin{figure}[tbh]
\includegraphics[width=\linewidth]{{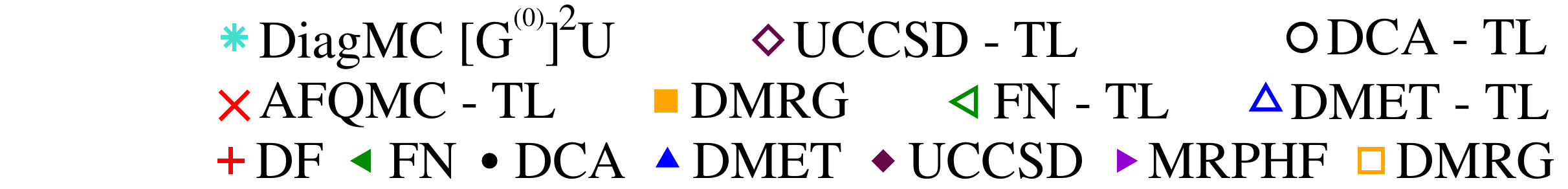}}
\includegraphics[width=\linewidth]{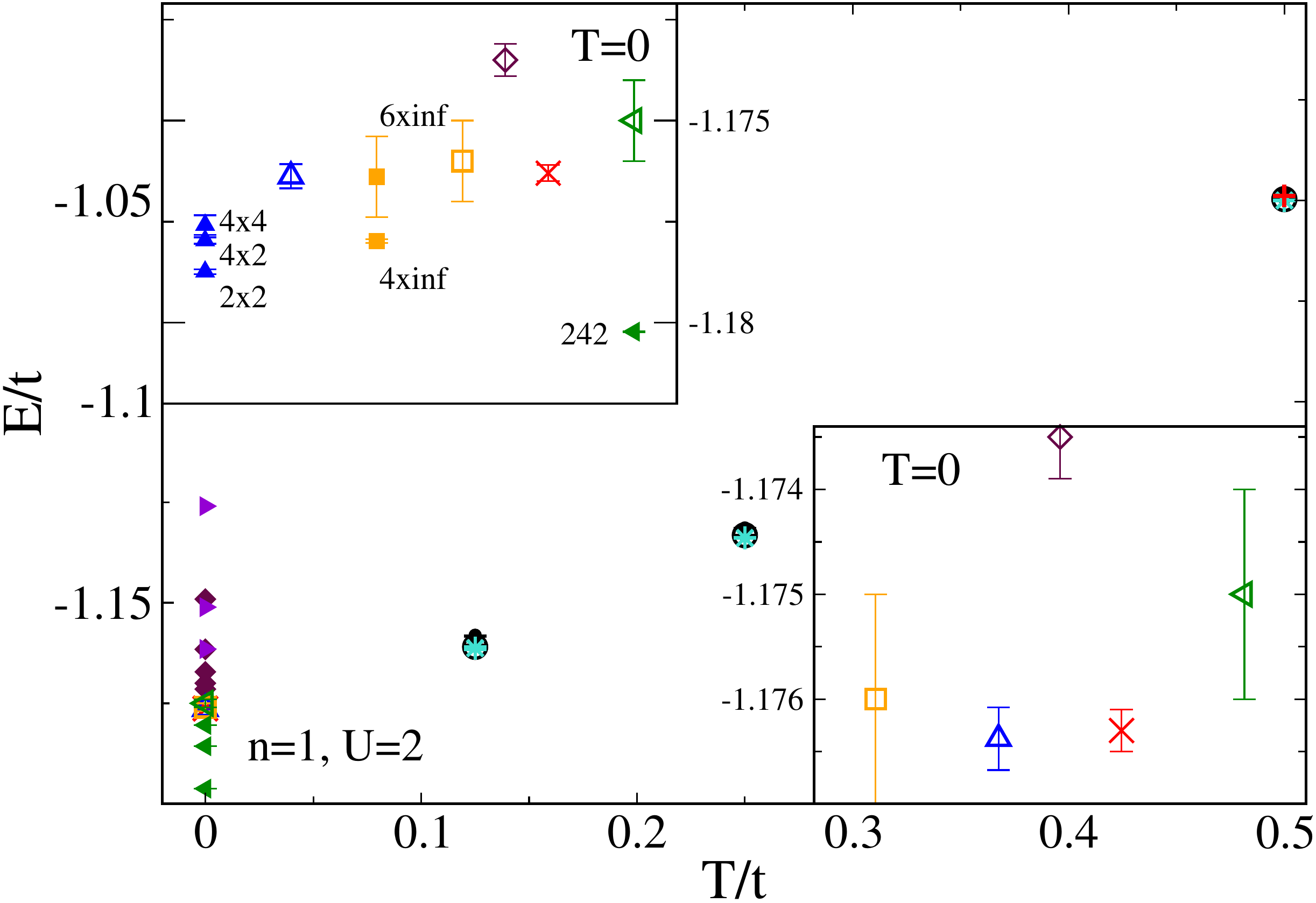}
\caption{\label{fig:U2_n1_E} (Color online)  Data for $n=1$ for $U/t=2$. Main panel: temperature dependence of $E/t$ compiled from various techniques.  Solid symbols represent finite systems, open symbols represent extrapolations to the thermodynamic limit (TL).  Finite $T$ results are shown for DCA (black circles) and DiagMC (blue asterisks), and  zero-$T$ data from AFQMC (red crosses), DMET (blue triangles), UCCSD (maroon diamonds), MRPHF (purple triangles), DMRG (orange squares), and FN (green triangles).
Top left inset: zoom in to the zero-$T$  data from UCCSD, DMET, FN, DMRG, and AFQMC, and DMET. Bottom right panel: enlarged region of the top left inset showing thermodynamic limit data for DMRG, DMET, UCCSD, FN and AFQMC data at $T=0$, including error bars, extrapolated to the infinite system size.  }
\end{figure}
\begin{figure}[tbh]
\includegraphics[width=\linewidth]{{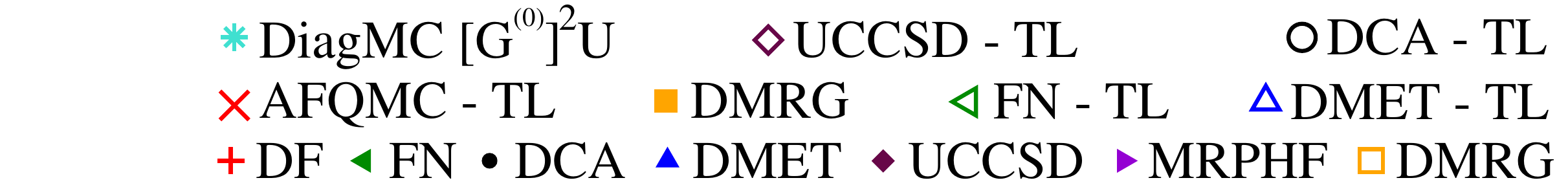}}
\includegraphics[width=\linewidth]{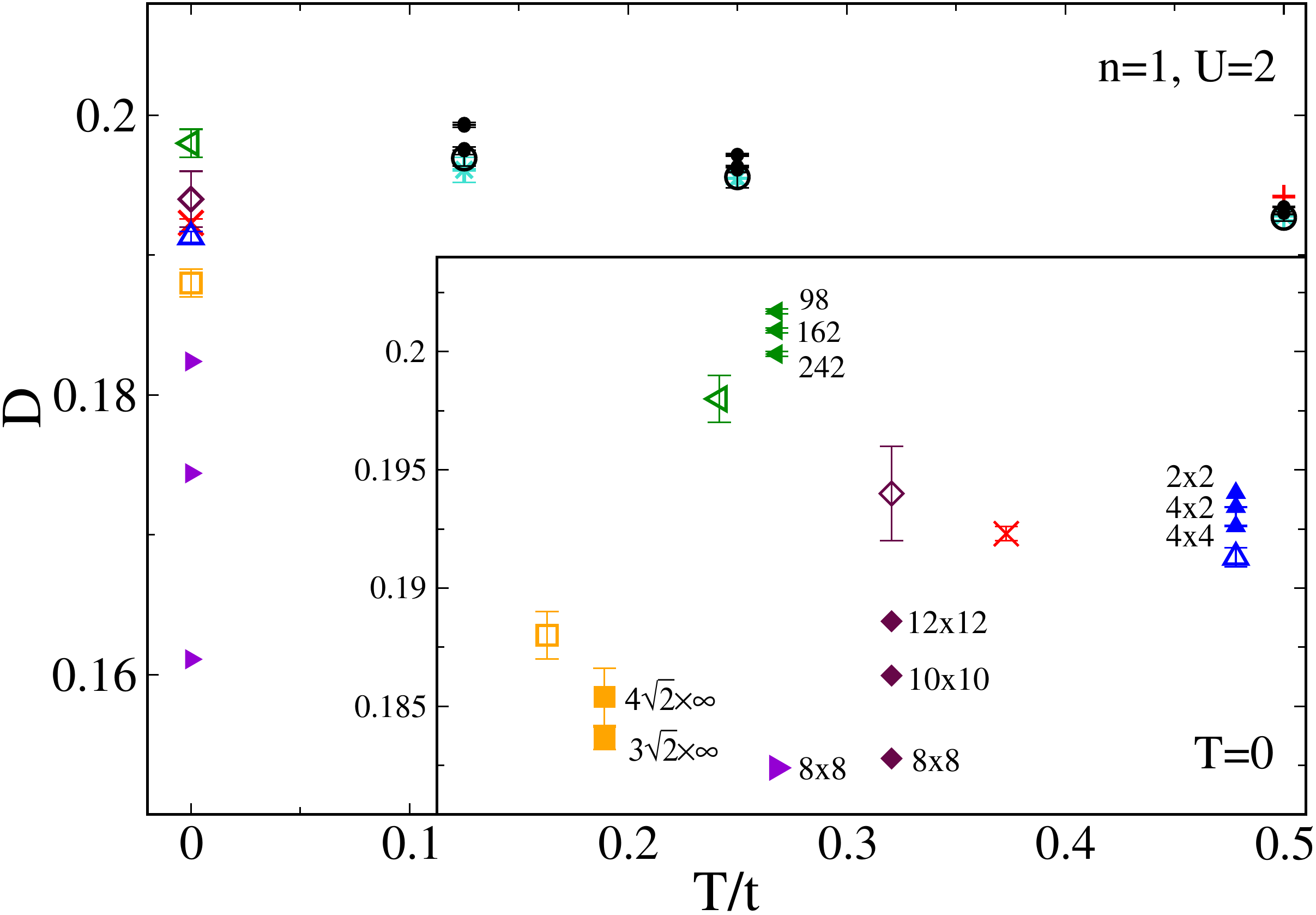}
\caption{\label{fig:U2_n1_D} Data for $n=1$ for $U/t=2$. Main panel: temperature dependence of double occupancy, $D$, compiled from various techniques.  Solid symbols represent finite systems, open symbols represent extrapolations to the thermodynamic limit (TL).  Finite $T$ results are shown for  DCA (black circles) and DiagMC (turquoise asterisks), and  zero-$T$ data from DMET (blue triangles), UCCSD (maroon diamonds), MRPHF (purple triangles), DMRG (orange squares), and FN (green triangles).
Inset: zoom in to the zero-$T$  data from  FN, UCCSD, MRPHF, DRMG, DMET, and AFQMC.}
\end{figure}

In Figs.~\ref{fig:U2_n1_E} and \ref{fig:U2_n1_D} we present results for $U/t=2$ and $n=1$, the half-filled weak coupling regime. This regime is particularly easy for methods based on a weak coupling expansion around a non-interacting system, and many of the algorithms show uncertainties that are much smaller than in the intermediate or strong interaction limit. 

At non-zero $T$, the data from two types of DiagMC and from DCA in the thermodynamic limit agree within uncertainty.
The values smoothly connect to the $T=0$ values, except for MRPHF energies, which are higher than the ground state energies obtained by the other methods and higher than the energies obtained for the lowest temperature point obtained from both finite-T methods.  

At $T/t=0.5$ we show a result from DF.  
As was the case in the strong coupling regime, the DF procedure produces an energy estimate consistent with DCA results.  In this case, with only weak finite size dependence, the underlying DMFT approximation differs from DCA by only 0.4\%.  The DF value improves on the DMFT and  differs from the extrapolated DCA results by only 0.07\%.

In the lower right inset of Fig.~\ref{fig:U2_n1_E}, we present $T=0$ extrapolations. As in the case of larger interactions,  DMET and AFQMC (which is numerically exact in this situation) agree
precisely while FN is slightly higher in energy but compatible within two error bars. 
In the upper left inset we explore the finite size effects of the methods. In the case of DMET, these finite size effects are small, and can be extrapolated with small error bars. FN shows much larger finite size effects, approaching the thermodynamic limit energy from below (only the largest system size is visible on the scale of the plot).  DMRG results with phase averaging are precise even at $U/t=2$, though much larger uncertainties than at $U/t=8$ are present.  

The results of MRPHF, outside of the scale shown by the inset, show a gradual decrease of the energy with increasing system size: $4\times 4$ yields $E/t=-1.1260$, $6\times 6$ $E/t=-1.1515$, and $8\times 8$ $E/t=-1.1629$. 
UCCSD results, here shown for cluster sizes of $12 \times 12$, $10 \times 10$, $8\times8$, and $6\times 6$, show large finite size effects.  With the aid of an extrapolation the value in the thermodynamic limit is estimated higher than other techniques. The deviation of the thermodynamic limit value from  AFQMC and DMET is on the order of $2\times 10^{-3}t$, suggesting that excitations beyond the singles and doubles level are important even in this relatively weak coupling regime.

Finally, Fig.~\ref{fig:U2_n1_D} shows the double occupancy for these parameters. We see an increase in finite size effects in DCA as we progress to lower temperature.  Reasonable agreement with DiagMC is achieved in the double occupancy. At temperatures lower than our lowest temperature, the double occupancy will need to dip, as was the case at $U/t=4$, in order for the finite $T$ data to be consistent with $T=0$.
Similar to the strong coupling case, DF provides only a slight shift to the double occupancy, a minimal improvement over DMFT alone.  

The ground-state double occupancies are very precisely determined by AFQMC and DMET which are in agreement. The FN value is somewhat overestimated.  Results from DMRG fall below AFQMC and DMET
and the error bar underestimates the uncertainty. The larger error appears consistent
with the difficulty in treating the small $U$ limit in the DMRG calculations. The results from MRPHF show an improvement in $D$ as the system size is increased, consistent with the behavior for the energy.  In the case of UCCSD, since it is an expansion in the coupling strength, at weak coupling the procedure is more reliable, and while there are substantial finite size effects, the extrapolation produces a result within error bars of AFQMC, and in general agreement with DMET.

\section{Frequency and Momentum Dependence}\label{results_kw_plots}

\begin{figure}
\includegraphics[width=\linewidth]{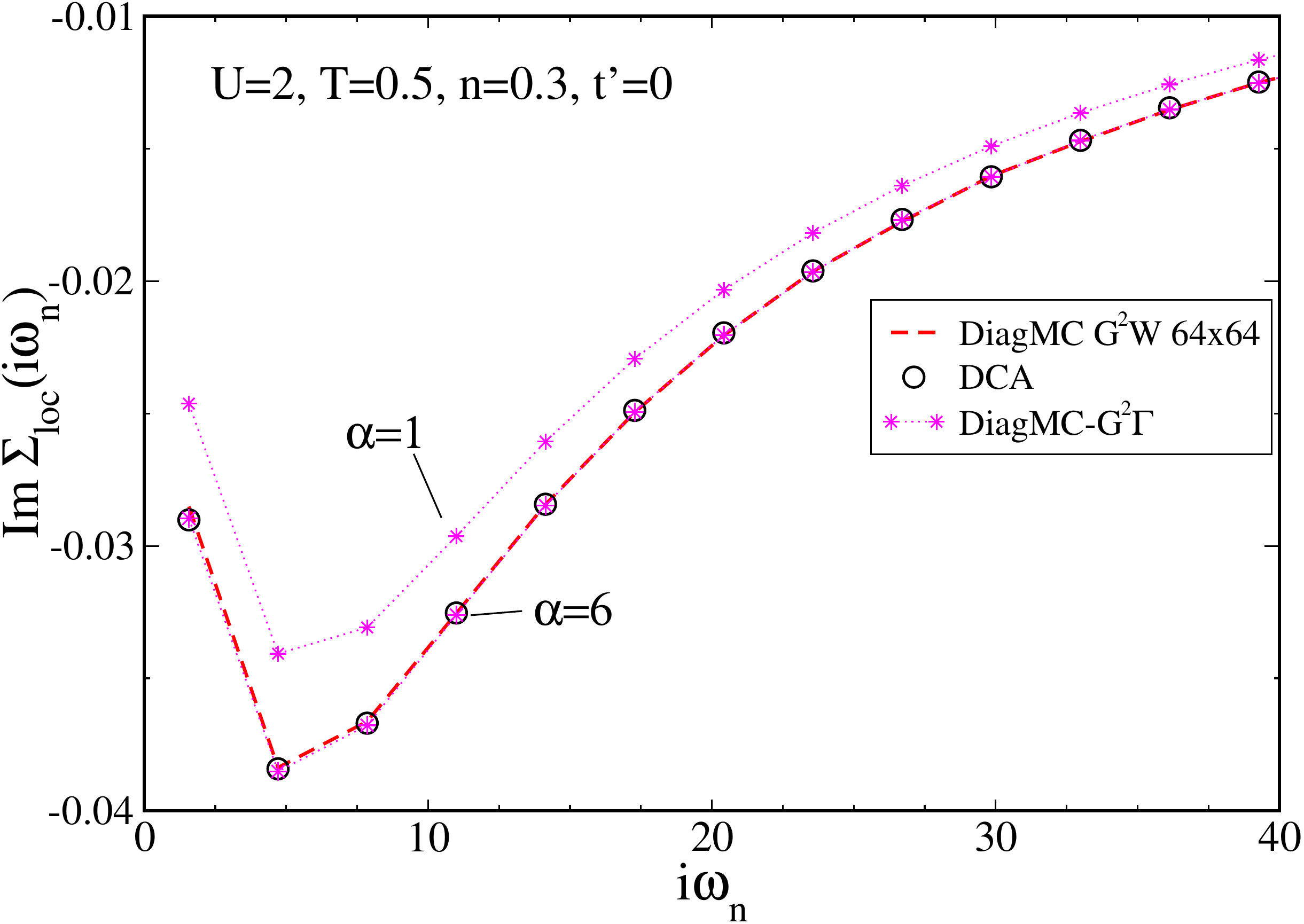}
\caption{\label{fig:freqdep} (Color online)  Imaginary part of the local self energy, ${\rm Im}\Sigma_{loc}(i\omega_n)$ at $U/t=2$, $T/t=0.5$ and $n=0.3$ from DCA and DiagMC.  In the case of DiagMC-$G^2\Gamma$ we label $\alpha$, the series order from Eqn~(\ref{eqn:diagmc}).   }
\end{figure}
\begin{figure}
\includegraphics[width=\linewidth]{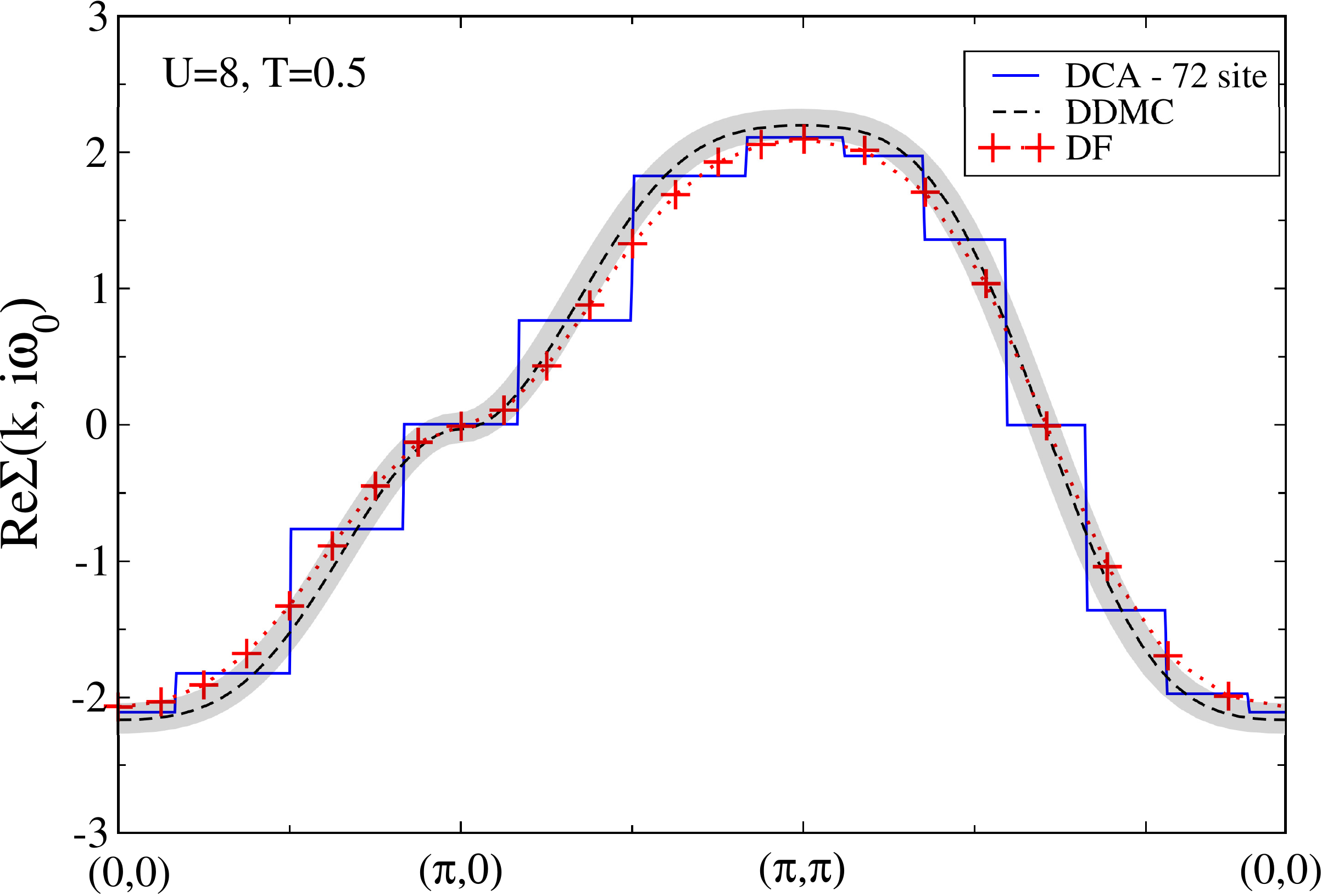}
\caption{\label{fig:kdep} (Color online)  Comparison of the lowest Matsubara frequency, $i\omega_0$, real part of the self energy, ${\rm Re}\Sigma(k, i\omega_0)$, obtained from DF (red) compared to 72-site DCA calculations (black) plotted as a function of momentum, $k$, throughout the Brillouin zone for $n=1.0$, $U/t=8$, $T/t=0.5$. The Dual Fermion and DCA self-energies are plotted as step functions. Also included are interpolated results obtained by diagrammatic determinantal Monte Carlo\cite{rubtsov:2005,Burovski06, Kozik13, kozik:2015} (dashed black) with a gray shading to indicate the level of uncertainty. }
\end{figure}
\begin{figure}
\includegraphics[width=\linewidth]{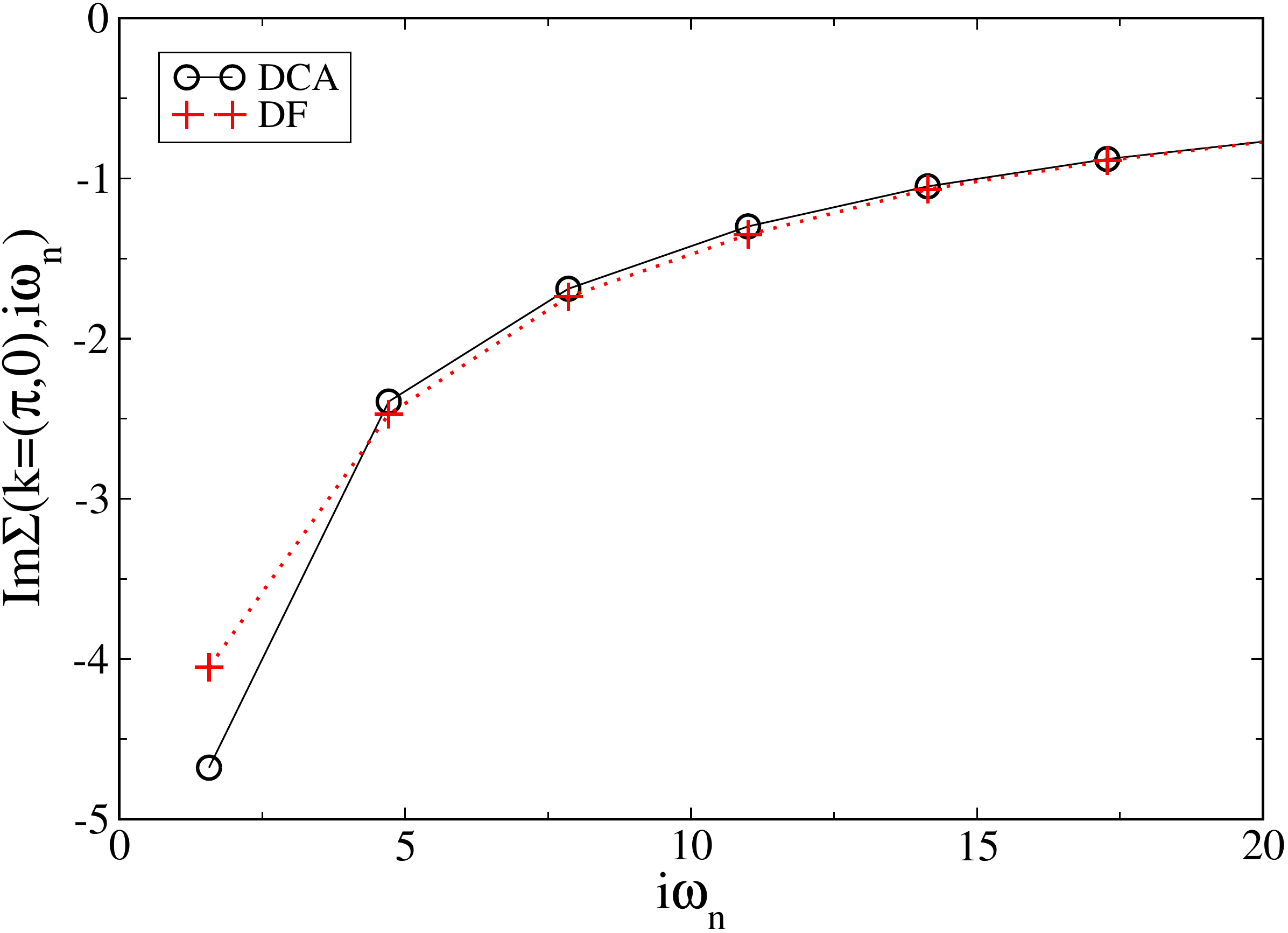}
\caption{\label{fig:imag_sigma_freqdep} (Color online)  Comparison of the frequency dependence of the imaginary part of the self energy, ${\rm Im}\Sigma(k, i\omega_n)$, at fixed $k=(\pi,0)$ obtained from DF (red) compared to 72-site DCA calculations (black) plotted for $n=1.0$, $U/t=8$, $T/t=0.5$.  Also shown are results from dynamical vertex approximation, $D\Gamma A$, (blue).\cite{toschi:2007,schaefer:2015} }
\end{figure}

Next we discuss single-particle finite temperature properties. All finite-temperature algorithms discussed in this work are based on  approximations of the single-particle self-energy. We show three characteristic plots for this quantity: Figure~\ref{fig:freqdep} shows the imaginary part of the local self-energy as a function of Matsubara frequency, Figure~\ref{fig:kdep} shows the dependence of the real part of the lowest Matsubara frequency on k-space, and Figure~\ref{fig:imag_sigma_freqdep} shows the frequency-dependence of the imaginary part of the self-energy for a specific momentum. Any discrepancy in the energy or double occupancy is the result of discrepancies in the single-particle self-energy.

The data shown in Fig.~\ref{fig:freqdep} is obtained for weak interaction strength $U/t=2$ and for a density $n=0.3$.
In this metallic regime, self-energies are small. Black circles denote the imaginary part of the local self energy from an $N_c=20$ DCA calculation, which for these parameters shows essentially no finite size effects. The data agrees perfectly with DiagMC-$G^2W$ data shown as red dashed lines, and convergence of the DiagMC-$G^2\Gamma$ method to the result of the other two methods (stars, magenta dotted line) is observed as a function of expansion order $\alpha$. 
This agreement implies that the local physics is captured well by all three algorithms.
 
In Fig.~\ref{fig:kdep} we  examine momentum dependent data. We show a path of $(k_x, k_y)$ through the Brillouin zone and plot the real part of the self energy at the lowest Matsubara frequency at $U/t=8$, $\beta=2$ and $n=1$. DiagMC data is not available in this regime, but for comparison we plot large DCA cluster results ($N_c=72$) and results from continuous-time lattice Monte Carlo simulations (DDMC, see Refs.~\onlinecite{rubtsov:2005,Burovski06, Kozik13, kozik:2015}). The DCA approximation (blue lines) produces a step-discretized self-energy which is in approximate agreement with the momentum dependence from other techniques.  Discrepancies between the approximate DF method and the (essentially converged) DDMC method are visible but within the uncertainty of the DDMC comparison data.  Any discrepancies are expected to  rapidly disappear at higher Matsubara frequencies, which can be seen in a comparison of the imaginary part of the self-energy (Fig.~\ref{fig:imag_sigma_freqdep}) of DF and DCA at fixed momentum $k=(\pi,0)$.  For comparison/verification purposes, we include results from dynamical vertex approximation ($D\Gamma A$, see Refs.~\onlinecite{toschi:2007,schaefer:2015, schaefer:2015:arxiv}) which, in a spirit similar to DF solves the model in an expansion of two-particle vertex functions.  We find that the results from DF and $D\Gamma A$ are consistent.

\section{Tables for Ground State Properties}
We conclude the discussion of our results with a list of the thermodynamic limit estimates for the half-filled $t'=0$ case. Table \ref{table:zeroTE} shows a list of energies in the thermodynamic limit, obtained from AFQMC, DMRG, DMET, and FN. The MRPHF and UCCSD values presented show the value for the largest finite size system studied. The uncertainties presented are the best uncertainties available within each algorithm and do not contain an assessment of systematic  errors (e.g. fixed node or truncation in expansion order). We see that for much of phase space, errors are a few times $10^{-4}t$ and values between the techniques are remarkably consistent.

Table \ref{table:zeroTD} shows the double occupancy for the same values. Relative errors for the double occupancy are of the same magnitude as for the total energy, and values in the thermodynamic limit are consistent within error bars. 
Tables \ref{table:zeroTnp8} and \ref{table:zeroTnp875} present ground state energies for the densities $n=0.8$ and $0.875$ for values of $U/t$ considered in this work.
The full table of values for the data presented in this paper is available online as supplemental material to this paper.\cite{suppl}

Not all quantities are as consistent as the energies.  This is especially true for order parameters and correlation functions, where discrepancies outside of error bars are present. Presumably, many competing phases exist in a narrow energy window near the ground state, and the most favorable state found in each method will depend on details of the finite size system and the approximation. Table \ref{table:m_compare} shows the comparison between the magnetization that DMET observes and the magnetization found in AFQMC for the full range of $U/t$ at half filling. 
At weak interaction strength, DMET finds a larger polarization than AFQMC even though the energies agree to all significant digits, as a result of DMET finite size scaling from small clusters.  At large interaction strength, AFQMC gives a polarization with large statistical fluctuation despite very accurate energies.
 Similar behavior (not shown here) is apparent for other variables, {\it e.g.} the $d$-wave order parameter or the stripe geometry observed at $U/t=8$ and $n=0.875$.
\begin{table*}[bth]
\centering
\begin{tabular}{|l|cc||cc||cc||cc||cc|}
\hline
    \multicolumn{1}{|c||}{U} & \multicolumn{2}{c||}{2}  & \multicolumn{2}{c||}{4}& \multicolumn{2}{c||}{6}& \multicolumn{2}{c||}{8}& \multicolumn{2}{c|}{12} \\
     \hline
    \hline
    AFQMC     & -1.1763    & 0.0002     & -0.8603    & 0.0002     &    -0.6568    & 0.0003    & -0.5247    & 0.0002    & -0.3693     & 0.0002    \\
    \hline
    DMET       & -1.1764    & 0.0003    & -0.8604     & 0.0003  &   -0.6562     & 0.0005    &   -0.5234  & 0.0010    &   -0.3685  & 0.0010     \\
    \hline
    DMRG       & -1.176        & 0.001       & -0.8605    & 0.0005    & -0.6565          & 0.0001      & -0.5241    & 0.0001    & -0.3689    & 0.0001   \\
\hline
    FN             & -1.175     & 0.001    & -0.8575        & 0.0003    & -0.6551       & 0.0001    & -0.52315     & 0.00005    & -0.36835    & 0.00005    \\    
    \hline
     MRPHF    & -1.1628   & $[8\times8]$          & -0.8554         & $[8\times8]$           & -0.6512       & $[8\times8]$ & -0.5169   &$[8\times8]$            & -0.3626    & $[8\times8]$     \\
    \hline
    UCCSD             & -1.1735     & 0.0004    & -0.8546        & $[14 \times 14]$   & -0.6510       & $[10\times 10]$    & -0.5191    &  $[10\times 10]$       &  -0.3647  &  $[10\times 10]$   \\
    \hline
    UCCSDTQ*             & -1.1749     & --    & -0.8610        & --   & -0.6582       & --    & -0.5255    &  --       &  -0.3696  &  --   \\
    \hline
   \end{tabular}
\caption{\label{table:zeroTE}Zero-temperature energy and uncertainty for $n=1$, $T=0$, for a range of interaction strengths $U$, obtained from AFQMC, DMET, DMRG, FN, MRPHF and UCCSD. Where extrapolations to the TL are not available, finite size geometries are listed in lieu of uncertainties. UCCSDTQ* data estimates higher order corrections by including triples from a $[6\times6]$ and quadruples from a $[4\times4]$.  UCCSD data for $U/t > 4$ provides nearly
converged energy estimates with respect to system size.}
\end{table*}

\begin{table*}[bth]
\centering
\begin{tabular}{|l||cc||cc||cc||cc||cc|}
\hline
    \multicolumn{1}{|c||}{U} & \multicolumn{2}{c||}{2}  & \multicolumn{2}{c||}{4}& \multicolumn{2}{c||}{6}& \multicolumn{2}{c||}{8}& \multicolumn{2}{c|}{12} \\
     \hline
    \hline
    AFQMC     &   0.1923     & 0.0003 &0.1262  & 0.0002      &  0.0810  & 0.0001         & 0.0540  & 0.0001            & 0.0278    & 0.0001    \\
    \hline
    DMET       &  0.1913      & 0.0004 &  0.1261  & 0.0001 &   0.08095    & 0.00004 &    0.05398    & 0.00007    &   0.02780    & 0.00003    \\
    \hline
    DMRG      &   0.188  & 0.001        &    0.126  & 0.001       &   0.0809  & 0.0003           &    0.0539   &  0.0001       &   0.0278  & 0.0001 \\
    \hline
FN            & 0.198  & 0.001         &   0.125  & 0.001           &  0.0803  & 0.0002             &0.0535  & 0.0001       &   0.0278  & 0.0002  \\
    \hline    
     MRPHF    &  0.1824  & $[8\times8]$           &   0.1262  & $[8\times8]$            &   0.0818  & $[8\times8]$               &  0.0544  &$[8\times8]$                     &   0.0275  & $[8\times8]$  \\
    \hline
          UCCSD             & 0.194     & 0.002    & 0.1268        & $[12\times12]$   & 0.0807       &$[10\times10]$    & 0.0537    & $[10\times10]$    & 0.0276     & $[10\times10]$ \\
    \hline
   \end{tabular}
\caption{\label{table:zeroTD}Zero-temperature double occupancy and uncertainty for $n=1$, for a range of interaction strengths $U$, obtained from AFQMC, DMET, DMRG, FN, MRPHF, and UCCSD. Where extrapolations to the TL are not available, finite size geometries are listed in lieu of uncertainties.}
\end{table*}

\begin{table*}[bth]
\centering
\begin{tabular}{|l||cc||cc||cc||cc|}
\hline
\multicolumn{9}{|c|}{n=0.8}\\
\hline
    \multicolumn{1}{|c||}{U} & \multicolumn{2}{c||}{2}  & \multicolumn{2}{c||}{4}& \multicolumn{2}{c||}{6}& \multicolumn{2}{c|}{8} \\
     \hline
    \hline
    AFQMC     & -1.306    &  0.002     & -1.110     & 0.003    &    --   &--   & --    & --      \\
    \hline
    DMET       & -1.3062    & 0.0004    & -1.108     & 0.002  &   -0.977     & 0.004    &  -0.88   & 0.03   \\
    \hline
    DMRG       & --       & --       & -1.104     & 0.0014   & --          & --      & --    & --      \\
\hline
    FN             & -1.3044      & 0.0007   & -1.1032         & 0.0007    & -0.967       & 0.001    &-0.877     & 0.001       \\    
    \hline
     MRPHF    & -1.2931    & $[10\times4]$          & -1.0910          & $[10\times4]$             & -0.9454        & $[10\times4]$  & -0.8415    & $[10\times4]$               \\
    \hline
    UCCSD             & -1.3065      & $[10\times16]$     & -1.0868        & $[10\times16]$    & -0.9300        & $[10\times16]$     & -0.8233     & $[10\times16]$      \\
    \hline
    UCCSDT*            & -1.3078      & $[10\times4]$     & -1.0981        & $[10\times4]$     & -0.9607        & $[10\times4]$      & -0.8641     & $[10\times4]$       \\
    \hline
   \end{tabular}
\caption{\label{table:zeroTnp8}Zero-temperature energy and uncertainty  for $n=0.8$, $T=0$, for a range of interaction strengths $U$, obtained from AFQMC (constrained path), DMET, DMRG, FN, MRPHF and UCCSD. Where extrapolations to the TL are not available, finite size geometries are listed in lieu of uncertainties. }
\end{table*}

\begin{table*}[bth]
\begin{threeparttable}
\centering
\begin{tabular}{|l||cc||cc||cc||cc|}
\hline
\multicolumn{9}{|c|}{n=0.875}\\
\hline
    \multicolumn{1}{|c||}{U} & \multicolumn{2}{c||}{2}  & \multicolumn{2}{c||}{4}& \multicolumn{2}{c||}{6}& \multicolumn{2}{c|}{8} \\
     \hline
    \hline
    AFQMC     & --   &  --     & -1.026      & 0.001    &    --   &--   &-0.766     & 0.001      \\
    \hline
    DMET       & -1.2721  &  0.0006    & -1.031    & 0.003  &   -0.863    & 0.013    &  -0.749 \tnote{a}    & 0.007    \\
    \hline
    DMRG & -- & --   & -1.028 &  $[6\times \infty]$&-- & --&  -0.759 & 0.004 \\
    \hline
    FN             & -1.270        & 0.002   & -1.0225         & 0.0015   &-0.854        & 0.002    & -0.749      & 0.002       \\    
    \hline
     MRPHF    & -1.2855    & $[16\times4]$            & -1.0195         & $[16\times4]$           & -0.8318        &  $[16\times4]$            &  -0.7094   & $[16\times4]$                \\
    \hline
    UCCSD             & -1.2667       & $[16\times12]$    & -1.0093         & $[16\times12]$   & -0.8298        & $[16\times12]$   & -0.7123     &$[16\times12]$    \\
    \hline
     UCCSDT*             & -1.2681       & $[16\times4]$    & -1.0253         & $[16\times4]$   & -0.8570        & $[16\times4]$  & -0.7434  & $[16\times4]$   \\
    \hline
   \end{tabular}
\caption{\label{table:zeroTnp875}Zero-temperature energy and uncertainty  for $n=0.875$, $T=0$, for a range of interaction strengths $U$, obtained from AFQMC (constrained path), DMET, DMRG, FN, MRPHF and UCCSD. Where extrapolations to the TL are not available, finite size geometries are listed in lieu of uncertainties. }
\begin{tablenotes}
            \item[a] Due to strong spatial inhomogeneity observed at this filling, the TL extrapolated number in the table is not meaningful, as different cluster shapes show different orders. Out of the clusters used here, the $8\times 2$ impurity cluster is probably the best estimate with $E/t$=-0.755 (+/- 0.007). An exhaustive study of the DMET cluster size dependence at this filling will be carried out in the future.
   \end{tablenotes}
\end{threeparttable}
\end{table*}

\begin{table*}[tbh]
\centering
\begin{tabular}{|l||cc||cc||cc||cc|}
 \hline
    \multicolumn{1}{|c||}{U} & \multicolumn{2}{c||}{2}  & \multicolumn{2}{c||}{4}& \multicolumn{2}{c||}{6}& \multicolumn{2}{c|}{8} \\
      \hline
     Technique & $m$ & $\delta m$& $m$ & $\delta m$ & $m$ & $\delta m$& $m$ & $\delta m$ \\
     \hline
   AFQMC     & 0.094  & 0.004    & 0.236    & 0.001     & 0.280    & 0.005     & 0.26    & 0.03        \\
     \hline
    DMET     & 0.133  & 0.005     & 0.252 & 0.009     &   0.299 & 0.012    &   0.318  & 0.013   \\
     \hline
   \end{tabular}
 \caption{\label{table:m_compare}Comparison magnetization data from DMET and AFQMC at $n=1$.}
\end{table*}

\section{Conclusions \label{conclusions}}

In this paper we have presented a detailed examination of results  for static and dynamic properties of the two dimensional Hubbard model at correlation strengths ranging from weak to intermediate to strong coupling, and at various carrier concentrations,  obtained using  state of the art numerical methods. We believe the results are useful for two reasons. First, the two dimensional Hubbard model is one of the paradigm models of quantum condensed matter theory, and it is therefore important to determine, as reliably as possible, the state of our knowledge about it. Second, solving the grand-challenge problem of determining the physics of interacting many-electron systems will require numerics, and as no one technique is likely to provide solutions in all regimes or for all quantities of physical interest, it is important to develop tools for assessing the strengths and weaknesses of different approaches.

We argue that the only quantities that can meaningfully be compared  between different approaches are estimates, with error bars, for the thermodynamic limit values of observables including local operators such as the energy, double occupancy, density (or chemical potential) and magnetization, as well as correlation functions such as the electron self-energy.  We restricted attention to methods and regimes for which large enough systems can be studied that reasonable extrapolations to the thermodynamic limit can be performed. Care is required in performing the extrapolations and we have found it useful to present both the extrapolated results and (in most cases) the finite size data that led to the extrapolation. 

Comparison of results obtained from different methods shows that the ground state properties of a substantial part of the Hubbard model phase space are now under numerical control (see e.g. Figs~\ref{fig:U8_n1_E_zeroT} and \ref{fig:U4_np8_E_Tzero}). Moreover, where there is agreement on the ground state properties, the non-zero temperature methods appear to connect smoothly to the ground state as the temperature is decreased, although a quantitative extrapolation to $T=0$ is not yet available.  The most substantial uncertainties exist at intermediate correlations (e.g. $U/t \approx 4 \to 8$) and at dopings near to but not equal to the half filling value $n=1$. In this intermediate coupling/near half filling regime several physically different states seem to have very similar energies, and small effects can favor the choice of one state over the other, leading to substantial uncertainties in physical quantities. Also, the issues associated with fermion sign problems seem to be most severe. Interestingly, it is this regime that is of most physical interest in connection with high-$T_c$ superconductivity in the copper-oxide materials. 

Where two or more methods produce results that agree within reasonable error bars, we take the result to be established, and appropriate for use as a benchmark. Tables of our benchmark results are contained in the supplemental material\cite{suppl} and made available online.  We expect that these results will be useful in validating new methods, or new implementations of existing methods. 

Turning now to prospects and open questions, we first observe that all of the methods we have considered to date have difficulty in the physically interesting intermediate coupling, near half filling regime. Development of new methods, or improvement of existing methods to deal with this regime is urgently needed. Further, we remark that our understanding of dynamical correlation functions, even ones as simple as the electron Green function, is much less advanced than our understanding of ground state properties and simple expectation values. Finally, we observe that the process of producing this paper, in particular the confrontation of each method with the body of related work produced by other methods, led in several cases to substantial improvements in algorithm and error analysis. We suggest that as quantitative numerics in the  many-electron field continues to evolve, intercomparison of different methods, leading to benchmarking on important model problems, will significantly advance the field.

\acknowledgments{We acknowledge the Simons Foundation for funding. The work at Rice University
was supported by NSF-CHE-1462434. GKC acknowledges funding from the US Department of Energy for the development of the DMET method through DE-SC0010530, and for its application to the Hubbard model and superconductivity through DE-SC0008624. F.B. and L.F.T. acknowledge support from PRIN 2010\_2010LLKJBX. SZ and HS acknowledge support from NSF (DMR-1409510) for AFQMC
method development. MQ was also supported by DOE (DE-SC0008627), AEA by DOE ER 46932. YD and XWL  acknowledge support from NSFC (11275185). The AFQMC
calculations were carried out at the Oak Ridge Leadership Computing Facility at the Oak Ridge
National Laboratory, and at the computational facilities at the College of William and Mary.
}


\bibliography{bib_shortened.bib}
\end{document}